\documentclass[12pt,a4paper]{article}

\usepackage[english]{babel}

\usepackage{cite}
\usepackage{indentfirst}
\usepackage{amsmath}
\usepackage{amssymb}
\usepackage[pdftex]{graphicx,color}
\usepackage{slashed}
\usepackage{amsmath}
\numberwithin{equation}{section}

\definecolor{green}{rgb}{0.1,0.5,0.0}

\usepackage{geometry}
\begin{document}

\hfill IPPP/20/12 

\begin{center}

  {\bf \LARGE Contribution of the Darwin operator to}

  \vspace{0.4cm}

  {\bf \LARGE non-leptonic decays of heavy quarks}

\vspace*{1.6cm}

{\large Alexander Lenz, Maria Laura Piscopo, and Aleksey V. Rusov}

\vspace*{0.4cm}

\textsl{%
Institute for Particle Physics Phenomenology, Durham University, \\
DH1 3LE Durham, United Kingdom \\[5mm]
}
\vspace*{0.8cm}

\textbf{\Large Abstract}\\[10pt]
\parbox[t]{0.9\textwidth}{
We compute the Darwin operator contribution ($1/m_b^3$ correction) 
to the width of the inclusive non-leptonic decay of a $B$ meson ($B^+$, $B_d$ or $B_s$), 
stemming from the quark flavour-changing transition $b \to q_1 \bar q_2 q_3$, 
where $q_1,q_2 = u, c$ and $q_3 = d, s$. 
The key ideas of the computation are the local expansion of the quark propagator 
in the external gluon field including terms with a covariant derivative of 
the gluon field strength tensor and the standard technique of 
the Heavy Quark Expansion (HQE). 
We confirm the previously known expressions of the $1/m_b^3$ contributions to the
semi-leptonic decay $b \to q_1 \ell \bar \nu_\ell$, with $\ell = e, \mu, \tau$ 
and of the $1/m_b^2$ contributions to the non-leptonic modes.
We find that this new term can give a sizeable correction of about $- 4 \, \%$ 
to the non-leptonic decay width of a $B$ meson. For $B_d$ and $B_s$ mesons this turns
out to be the dominant correction to the free b-quark decay, while for the $B^+$ meson
the Darwin term gives the second most important correction - roughly 1/2 to 1/3 of the
phase space enhanced Pauli interference contribution.
Due to the tiny experimental uncertainties in lifetime measurements the incorporation of
the Darwin term contribution is crucial 
for precision tests of the Standard Model.}
\end{center}

\vspace*{3cm}
\newpage

\section{Introduction}
\begin{figure}[t]
\centering
\includegraphics[scale=0.36]{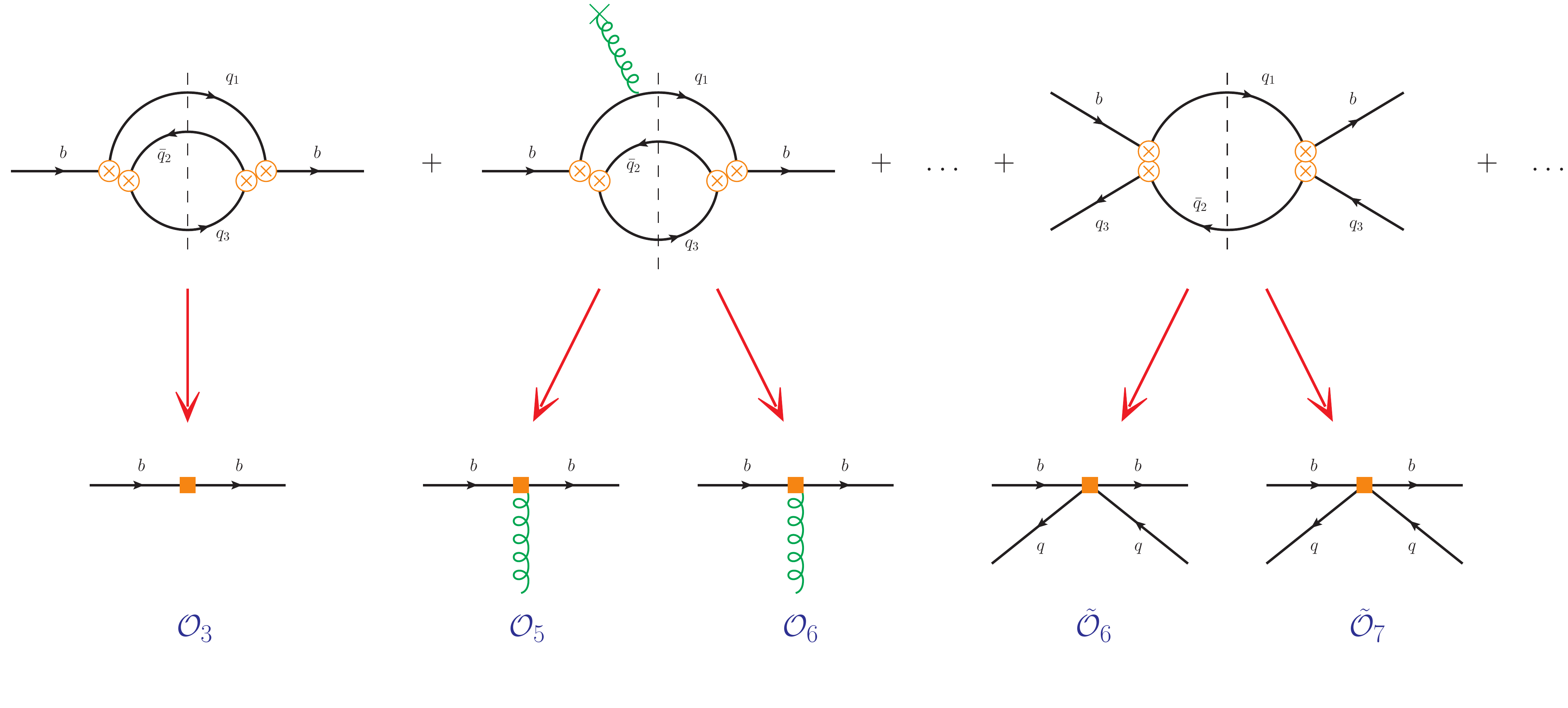}
\caption{The diagrams describing contributions to the HQE in Eq.~(\ref{eq:HQE}). 
The crossed circles denote the $\Delta  B =1$ operators $Q_i$ 
of the effective Hamiltonian while the squares denote the local $\Delta B = 0$ operators ${\cal O}_i$
and $\tilde{{\cal O}_i}$. The two-loop and the phase space enhanced one-loop diagrams correspond respectively
to the two-quark operators ${\cal O}_i$ and to the four-quark operators 
$\tilde{{\cal O}_i}$ in the HQE.}
\label{fig:intro}
\end{figure}

The total decay rate of heavy hadrons can be described by the Heavy Quark Expansion (HQE),
whose history goes back to the work of Shifman and Voloshin in the 1980ies
\cite{Khoze:1983yp,Shifman:1986mx}\footnote{For a more profound history see Ref.~\cite{Lenz:2015dra}.}, 
as the decay of the free heavy quark plus corrections
that are suppressed by inverse powers of the heavy quark mass. 
Since the $b$-quark mass is large compared to the typical hadronic scale,
the corrections are expected to be small and hence, back in 1986, the following $b$-hadron
lifetime ratios were expected \cite{Shifman:1986mx}
\begin{equation}
  {\frac{\tau(B_s)}{\tau(B_d)}}\Biggl|^{\rm HQE \, \, 1986} \! \! \! \! \! \approx 1 \, ,
  \hspace{0.7cm}
 {\frac{\tau(B^+)}{\tau(B_d)}}\Biggl|^{\rm HQE \, \, 1986} \! \! \! \! \! \approx 1.1 \, ,
  \hspace{0.7cm}
 {\frac{\tau(\Lambda_b)}{\tau(B_d)}}\Biggl|^{\rm HQE \, \, 1986} \! \! \! \! \! \approx 1 \, ,
 \hspace{0.7cm}
 \label{ratio_86}
\end{equation}
which are in good agreement with the current experimental averages obtained by HFLAV
(see the webupdate of Ref.~\cite{Amhis:2019ckw})
\begin{equation}
 {\frac{\tau(B_s)}{\tau(B_d)}}\Biggl|^{\rm HFLAV \, \, 2020} 
 \hspace{-1.1cm} = 0.998 \pm 0.004  \, ,
 \hspace{0.5cm}
 {\frac{\tau(B^+)}{\tau(B_d)}}\Biggl|^{\rm HFLAV \, \, 2020} 
 \hspace{-1.1cm} = 1.076 \pm 0.004  \, ,
 \hspace{0.5cm}
 {\frac{\tau(\Lambda_b)}{\tau(B_d)}}\Biggl|^{\rm HFLAV \, \, 2020} 
 \hspace{-0.9cm} = 0.969 \pm 0.006 \, .
 \hspace{0.5cm}
 \label{ratio_HFLAV}
\end{equation}
There has also been considerably progress on the theory side.
The total width of a $B$ meson with mass~$m_B$ and four-momentum $p_B^\mu$ is given by 
\begin{equation}
\Gamma (B)  = 
\frac{1}{2 m_{B}} \sum_{X}  \int \limits_{\rm PS} (2 \pi)^4 \delta^{(4)}(p_B- p_X) \, \, |\langle X(p_X)| {\cal H}_{\rm eff} | B(p_B) \rangle |^2,
\label{eq:Gamma-B}
\end{equation}
where ${\cal H}_{\rm eff}$ represents the effective weak Hamiltonian \cite{Buchalla:1995vs} describing all possible $b$-quark decays.
$PS$~denotes the phase space integration and we have summed over all possible final states $X$ into which the $B$ meson can decay. 
Eq.~(\ref{eq:Gamma-B}) can be related, via the optical theorem, to the double insertion of the effective Hamiltonian
and within the HQE framework one obtains
\begin{equation}
\Gamma(B) = 
\Gamma_3  +
\Gamma_5 \frac{\langle {\cal O}_5 \rangle}{m_b^2} + 
\Gamma_6 \frac{\langle {\cal O}_6 \rangle}{m_b^3} + ...  
 + 16 \pi^2 
\left[ 
  \tilde{\Gamma}_6 \frac{\langle \tilde{\mathcal{O}}_6 \rangle}{m_b^3} 
+ \tilde{\Gamma}_7 \frac{\langle \tilde{\mathcal{O}}_7 \rangle}{m_b^4} + ... 
\right],
\label{eq:HQE}
\end{equation}
with the matrix elements of the $\Delta B = 0$ operators
$\langle {\cal O}_Y \rangle = \langle B(p_B)  | {\cal O}_Y|B (p_B) \rangle$.
The structure of Eq.~(\ref{eq:HQE}) is diagrammatically represented in Fig.~\ref{fig:intro}.
The matrix elements of the operators ${\cal O}_i$ and~$\tilde {\cal O}_i$, denoting respectively two- and
four-quark operators, are suppressed by $i-3$ powers of the heavy quark mass $m_b$. The coefficients
$\Gamma_{i}$ and $\tilde \Gamma_{i}$ encode the corresponding short distance contributions.
The leading term $\Gamma_3$ describes the free $b$-quark decay and does not contain any non-perturbative corrections -
up to ${\cal O}(1/m_b)$
the matrix element of the operator $ {\cal O}_3 = \bar{b} b$ is simply one with the appropriate normalisation.
${\cal O}_5$ refers to the dimension-five kinetic and chromo-magnetic operators, proportional to two covariant
derivatives of the $b$-quark field. ${\cal O}_6$~includes the dimension-six Darwin and spin-orbit operators
with three covariant derivatives. So far the dependence on the spectator quark is only due to  different
values of the matrix elements and the coefficients $\Gamma_{3,5,6}$ are independent of the quark content of
the $B$ meson. The four-quark operators are phase-space enhanced (as indicated by the factor $16 \pi^2$)
and first arise at order $1/m_b^3$. The possible topologies, specifically weak annihilation (WA), Pauli interference (PI)
and weak exchange (WE), see Fig.~\ref{fig:PI-and-WE},
imply that the short distance coefficients are now dependent
on the spectator quark. The dimension-seven operators  
$\tilde{\cal O}_7$ contain one covariant derivative compared to $\tilde{\cal O}_6$. 
Due to the larger phase space, it is typically expected that $\tilde{\cal O}_6$ gives the dominant contribution
to the lifetime ratios \cite{Khoze:1983yp,Shifman:1986mx,Uraltsev:1996ta,Neubert:1996we,Lenz:2015dra}. 
\\
Currently $\Gamma_3$ is known at NLO-QCD
\cite{Hokim:1983yt,Altarelli:1991dx,Voloshin:1994sn,Bagan:1994zd,Bagan:1995yf,Lenz:1997aa,Lenz:1998qp,Krinner:2013cja}
for non-leptonic decays. NNLO-QCD corrections have been computed for semi-leptonic decays
\cite{Czarnecki:1997hc,Czarnecki:1998kt,vanRitbergen:1999gs,Melnikov:2008qs,Pak:2008cp,Pak:2008qt,Dowling:2008ap,Bonciani:2008wf,Biswas:2009rb,Brucherseifer:2013cu} and for non-leptonic decays the massless case was determined in full QCD 
(i.e. no effective Hamiltonian was used) in Ref.~\cite{Czarnecki:2005vr}.
$\Gamma_5$ was determined at LO-QCD for both semi-leptonic and non-leptonic decays 
\cite{Bigi:1992su,Blok:1992hw,Blok:1992he,Bigi:1992ne} and we confirm these results. For the semi-leptonic modes even NLO-QCD corrections
are available \cite{Alberti:2013kxa,Mannel:2014xza,Mannel:2015jka}.
$\Gamma_6$~was first computed at LO-QCD in Ref.~\cite{Gremm:1996df} 
and recently the NLO-QCD corrections were determined in Ref.~\cite{Mannel:2019qel}, 
both for the semi-leptonic case only.
$\tilde{\Gamma}_6$ is known at NLO-QCD for $B$-meson lifetimes
\cite{Beneke:2002rj,Franco:2002fc}
and $D$ meson lifetimes \cite{Lenz:2013aua}, while $\tilde \Gamma_7$ is only known
at LO-QCD  \cite{Gabbiani:2004tp}.
\\
This work presents the first determination of $\Gamma_6$ for non-leptonic decays. 
An interesting subtlety of the computation is mixing between four- and two-quark operators. 
Namely, at dimension-six the renormalised one-loop matrix elements of the operators $\tilde {\cal O}_6$
contribute to the coefficient of the Darwin operator through the diagram 
in Fig.~\ref{fig:ME-O-T-1-2}, ensuring the cancellation of the infrared~(IR) divergences 
that otherwise would appear in $\Gamma_6$. This feature has been intensively discussed 
for semi-leptonic decays \cite{Novikov:1983gd,Benson:2003kp,Bigi:2005bh,Fael:2019umf} 
- also under the name of "intrinsic charm", see e.g.~Refs.~\cite{Breidenbach:2008ua,Bigi:2009ym}.
\\
Finally, for numerical analysis, the values of the non-perturbative matrix elements  $\langle {\cal O}_Y \rangle $ are needed.
$\langle {\cal O}_5 \rangle $ and $\langle {\cal O}_6 \rangle $ can be extracted from fits to the
semi-leptonic spectrum for the case of $B_d $ and $B^+$ mesons, see e.g. 
Ref.~\cite{Alberti:2014yda}.
In the literature one can also find lattice determinations \cite{Bazavov:2018omf,Gambino:2017vkx,Aoki:2003jf,Kronfeld:2000gk,Gimenez:1996av}
and sum rule estimates \cite{Ball:1993xv,Bigi:1994ga,Neubert:1996wm} for these parameters.
For the $SU(3)_F$ violating ratios
$\langle B_s | {\cal O}_{5,6} | B_s \rangle / \langle B_d | {\cal O}_{5,6} | B_d \rangle$ 
one can use the theory estimates from Ref.~\cite{Bigi:2011gf}. 
The matrix elements of the four-quark operators $\langle \tilde{\mathcal{O}}_6 \rangle$ have been determined
by  Heavy Quark Effective Theory (HQET) sum rules \cite{Kirk:2017juj}. Violations of  $SU(3)_F$ are expected to yield visible effects and a calculation of these corrections with HQET sum rules - following Ref.~\cite{King:2019lal} - is currently been performed \cite{King:2020}.
Corresponding lattice results for the matrix elements of the four-quark operators would be highly desirable.
\\
Taking all the currently known contributions into account we arrive at significant
 improvements compared to the pioneering work in 1986, and the 2019 status of lifetime predictions reads
\cite{Kirk:2017juj,Lenz:2015dra}
\begin{equation}
  {\frac{\tau(B_s)}{\tau(B_d)}}\Biggl|^{\rm HQE \, \, 2019}       \hspace{-0.9cm} = 1.0006 \pm 0.0025 \, ,
  \hspace{0.5cm}
{\frac{\tau(B^+)}{\tau(B_d)}}\Biggl|^{\rm HQE \, \, 2019}       \hspace{-0.9cm} = 1.082^{+0.022}_{-0.026} \, ,
 \hspace{0.5cm}
 {\frac{\tau(\Lambda_b)}{\tau(B_d)}}\Biggl|^{\rm HQE \, \, 2019} \hspace{-0.9cm} = 0.935 \pm 0.054 \, ,
 \hspace{0.5cm}
 \label{ratio_19}
\end{equation}
which constitutes an impressive  confirmation of the validity of the HQE.
\\
The main motivations for this work are:
\begin{itemize}
\item 
In the case of the $\tau(B_s)/\tau(B_d)$ lifetime ratio several very pronounced cancellations 
are arising \cite{Beneke:1996gn,Keum:1998fd,Lenz:2011ti,Lenz:2015dra,Kirk:2017juj} among the contributions of $\tilde{\cal O}_6$ and ${\cal O}_5$, which could in principle make this 
ratio sensitive to the contribution of ${\cal O}_6$ - if $SU(3)_F$ violating
corrections are large for this term, as indicated by Ref.~\cite{Bigi:2011gf}.

\item 
It was found (see e.g. Refs.~\cite{Gremm:1996df,Benson:2003kp,Dassinger:2006md,Mannel:2010wj,Bigi:2011gf,Mannel:2017jfk})
that the $1/m_b^3$ correction in semileptonic inclusive decays 
$B \to X_c \, \ell \bar \nu_\ell$
are of a similar size as the $1/m_b^2$ terms due to enhanced Wilson coefficients.
This could lead to a visible effect in $\tau (B_s)/\tau (B_d)$, in particular 
if all other contributions are cancelling to a high degree.

\item 
As indicated in Eq.~(\ref{ratio_HFLAV}) the experimental precision achieved so far is very high,
enabling thus precision tests and an even higher precision seems to be achievable, 
see e.g. the two most recent results from LHCb \cite{Aaij:2019vot} and ATLAS \cite{Aad:2020jfw},
which interestingly differ significantly. 
The theory precision should of course cope with these experimental advancements.

\item 
According to the above arguments the lifetime ratio $\tau(B_s) / \tau(B_d)$ 
might thus provide a unique opportunity to test directly higher orders in the HQE 
that are otherwise just noise compared to the leading, numerically  dominant contributions. 
In that respect this might also increase our insights on the assumptions of quark hadron duality,
severely questioned around 20 years ago - mostly due to a very low experimental value 
of the $\Lambda_b$ lifetime, which seemed to be in severe conflict with the HQE expectation
from Eq.~(\ref{ratio_86}):
the 2002 HFAG average gives e.g. $\tau (\Lambda_b)/\tau(B_d) = 0.798(34) $, more than $5 \sigma$ 
away from the current experimental value given by Eq.~(\ref{ratio_HFLAV}). 
This signal for a violation of quark hadron duality was a clear false alarm.
Moreover, the first measurement of the decay rate difference of neutral $B_s$ mesons, 
$\Delta \Gamma_s$, has excluded large duality violating effects, see e.g. the discussion
in Refs.~\cite{Lenz:2012mb,Lenz:2011zz}. 
Smaller effects of duality violation can still appear and their potential size can be constrained 
by comparing experiment and theory for as many HQE observable as possible with a high precision, 
see e.g. Ref.~\cite{Jubb:2016mvq}.

\item 
Assuming the validity of quark hadron duality the $\tau(B_s)/\tau(B_d)$ lifetime ratio 
can be used to constrain invisible or hardly detectable $B$-meson decays, like $B_s \to \tau \tau$, 
at the per mille level, see e.g. Refs.~\cite{Bobeth:2011st,Bobeth:2014rda} - 
see also Ref.~\cite{Cornella:2020aoq} for an alternative way 
to constrain the potential size of the $b s \tau \tau$ couplings.
\end{itemize}
We will not present updated lifetime ratio predictions in this work, but we will postpone a new numerical study until the
$SU(3)_F$ violation ratio of the Bag parameters is available, see Ref.~\cite{King:2020}.
\\
The paper is structured as follows:
in Section \ref{sec:out} we outline the main ingredients of the calculation:  in Section  \ref{sec:outline} we
explain how the double insertion of the effective Hamiltonian
and the subsequent expansion in the external soft gluon field are performed. Section \ref{sec:logs}
is devoted to describe mixing between two- and four-quark operators at order $1/m_b^3$, which guarantees the cancellation of the
IR divergences. Our results are presented
in Section \ref{sec:results}
and we conclude in Section \ref{sec:discussion}. The expansion of the quark propagator in the external gluon field is discussed in more detail in Appendix~\ref{sec:propagator}, supplement material to Section~\ref{sec:logs} and Section~\ref{sec:results} can be found in Appendix~\ref{App:Mixing} and Appendix~\ref{App:Xs}, respectively, and for completeness we also present the non-leptonic results for $\Gamma_5$ in Appendix~\ref{App:C3C5}.

\section{Outline of the calculation}
\label{sec:out}
\subsection{Contribution of two-quark operators up to order \boldmath $1/m_b^3$}
\label{sec:outline}
\begin{figure}[t]\centering
\includegraphics[scale=0.5]{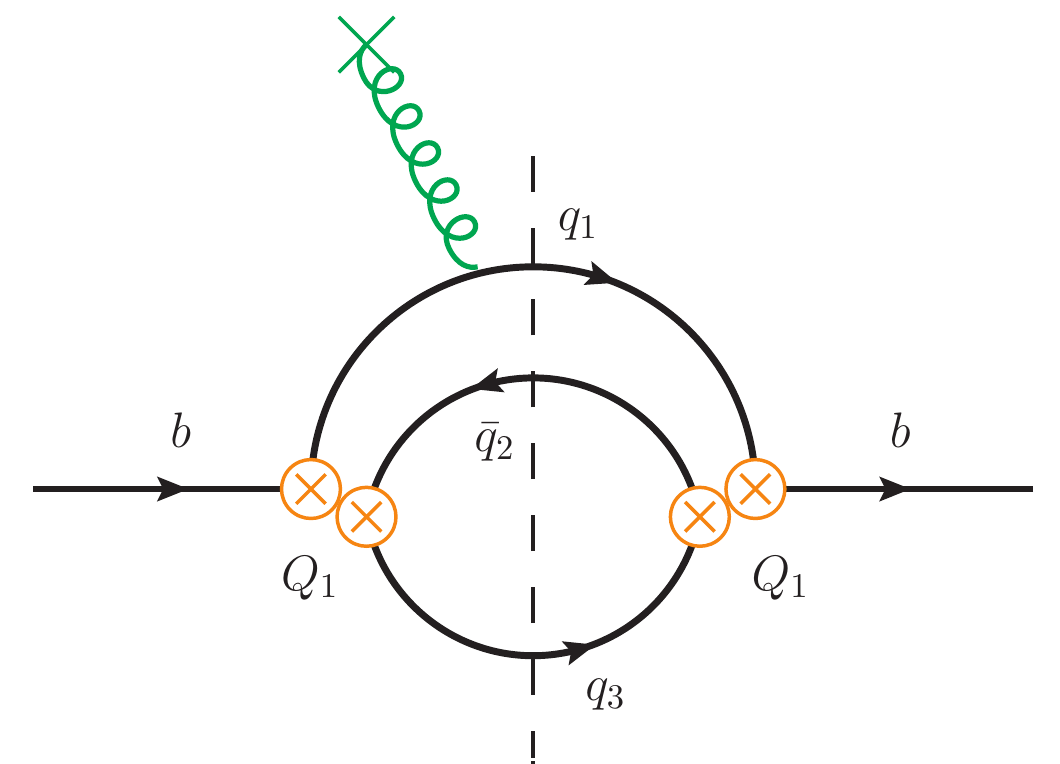}
\qquad
\includegraphics[scale=0.5]{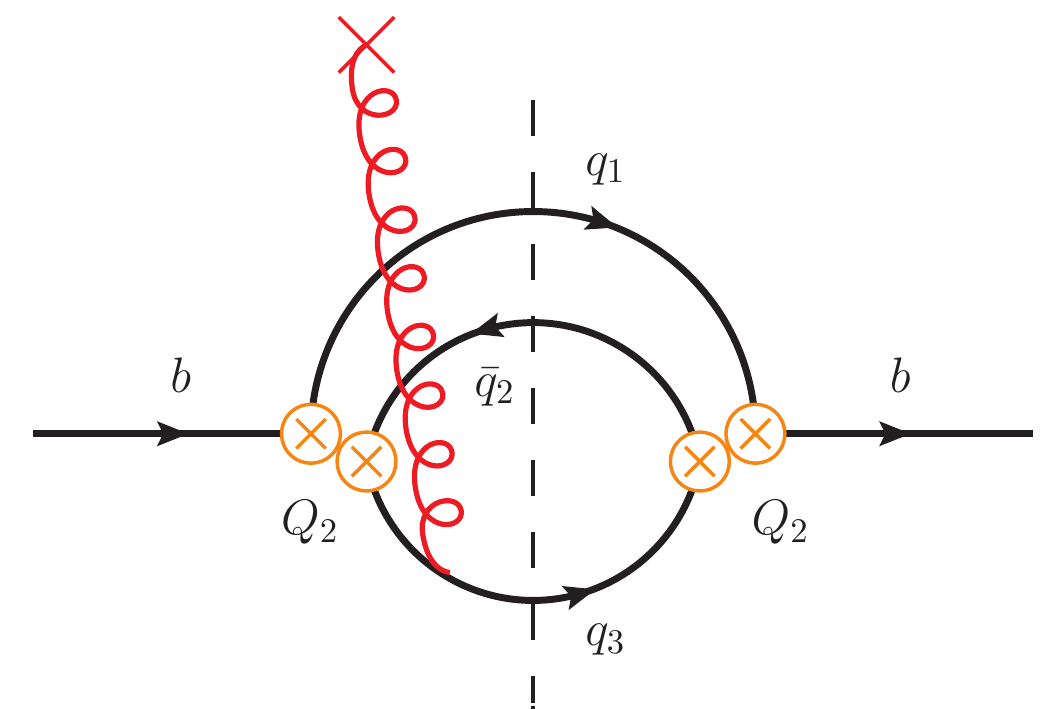}
\caption{The two-loop diagrams describing power corrections up to dimension-six from 
$Q_1 \otimes Q_1$ (left) and $Q_2 \otimes Q_2$ (right) contributions.}
\label{fig:O1-O1}
\end{figure}

\begin{figure}[t]\centering
\includegraphics[scale=0.5]{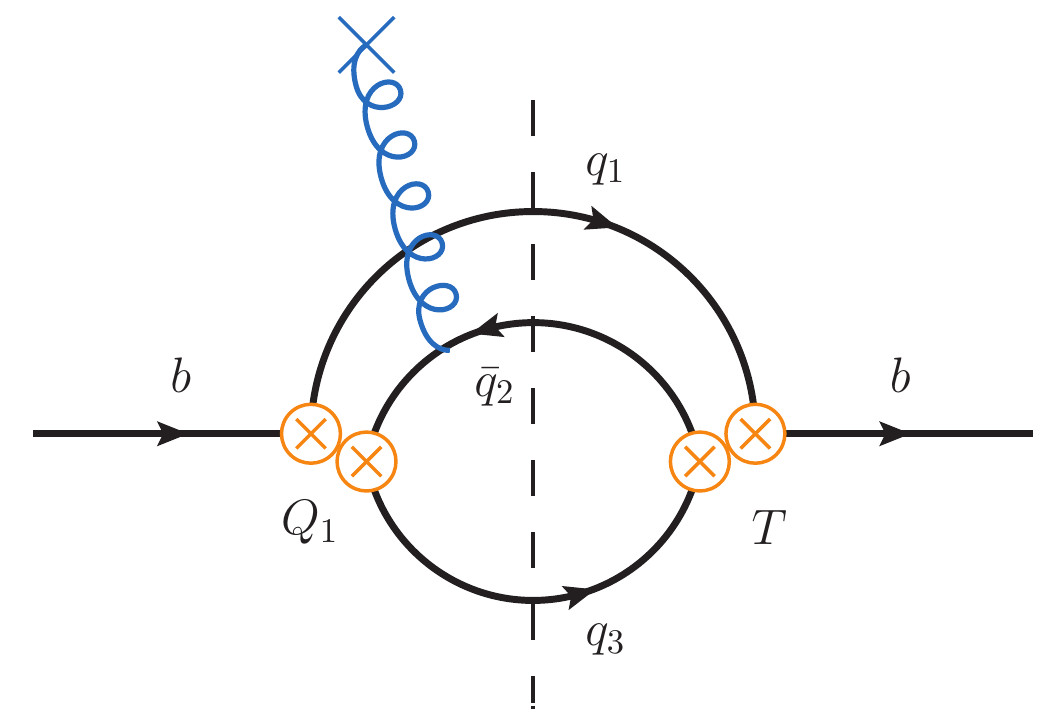}
\qquad
\includegraphics[scale=0.5]{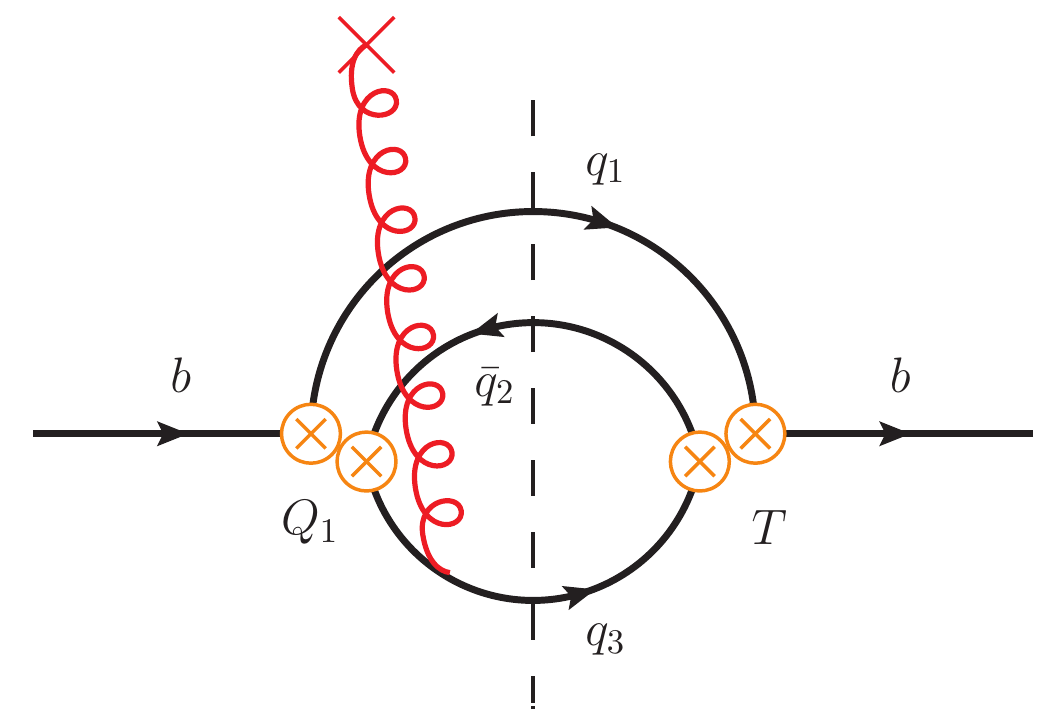}
\caption{The two-loop diagrams describing power corrections up to dimension-six from 
$Q_1 \otimes T$ contribution.}
\label{fig:O1-O2t}
\end{figure}

According to the optical theorem, the total width for the inclusive non-leptonic decay of a $B$ meson 
induced at the quark level by the flavour-changing transition $b \rightarrow q_1 \bar q_2 q_3$
(with $q_1,q_2 = u, c$ and $q_3 = d, s$),
can be computed from the discontinuity of the forward scattering matrix element: 
\begin{equation}
\Gamma_{\rm NL} (B)  = 
\frac{1}{2 m_{B}} \,  
{\rm Im} \, \langle B (p_B) | \,  i  \int d^4 x \,  T \left\{ 
{\cal L}_{\rm eff}(x) , {\cal L}_{\rm eff}(0) \right\}  |B (p_B) \rangle.
\label{eq:Gamma-def}
\end{equation}
The effective Lagrangian  ${\cal L}_{\rm eff} (x)$
reads
\begin{equation}
{\cal L}_{\rm eff} (x) = - \frac{4 G_F}{\sqrt 2} V_{q_1 b}^* V_{q_2  q_3}  
\left[ C_1 \,  Q_1 (x) + C_2 \, Q_2 (x) \right] + {\rm h.c.},
\label{eq:L-eff}
\end{equation}
where $G_F$ is the Fermi constant, $V_{q q^\prime}$ are the elements of the 
Cabibbo-Kobayashi-Maskawa (CKM) matrix \cite{Cabibbo:1963yz,Kobayashi:1973fv}, $C_{1,2}(\mu_1)$ denote the Wilson coefficients at 
the renormalisation scale $\mu_1 \sim m_b$ and $Q_{1,2}$ are the effective $\Delta B = 1$ four-quark operators 
\footnote{We consider only current-current operators with large Wilson coefficients.}
\begin{eqnarray}
Q_1 & = & (\bar q_1^i \Gamma_\mu b^i)(\bar q_3^{\, j} \Gamma^\mu q_2^{\, j}), 
\label{eq.O1-operator}\\
Q_2 & = & (\bar q_1^i \Gamma_\mu b^{\, j})(\bar q_3^{\, j} \Gamma^\mu q_2^i), 
\label{eq.O2-operator}
\end{eqnarray}
with $i,j$ standing for colour indices and $\Gamma_\mu \equiv  \gamma_\mu (1 - \gamma_5)/2$.
In our notation, $Q_1$ is the colour singlet operator and $Q_2$ the colour rearranged one,
opposite to  e.g. the notation in Ref.~\cite{Buchalla:1995vs}.
Three combinations of operators enter the decay rate in Eq.~(\ref{eq:Gamma-def}):
$Q_1 \otimes Q_1$, $Q_2 \otimes Q_2$ and $Q_1 \otimes Q_2$. We use the completeness
property of the colour matrices $t^a_{ij}$\footnote{The colour matrices satisfy the following relations: Tr$[t^a] = 0$ and
Tr$[t^a,t^b] = (1/2) \, \delta^{ab}$.}
\begin{equation}
t^a_{ij}t^a_{lm} = 
\frac{1}{2} \left( 
\delta_{im} \delta_{jl} - \frac{1}{N_c} \delta_{ij}\delta_{lm}
\right),
\label{eq:colour-trafo}
\end{equation}
to rewrite
\begin{equation}
Q_1 \otimes Q_2 = \frac{1}{N_c} (Q_1 \otimes Q_1) + 
2 \, (Q_1 \otimes T),
\label{eq.O1O2operator}
\end{equation}
where the colour octet operator $T$ is given by
\begin{equation}
T =  
(\bar q_1^{\, i} \Gamma_\mu \, t^a_{ij} \, b^j)(\bar q_3^{\, l} \, \Gamma^\mu \, t^a_{lm} \, q_2^m).
\label{eq.octet-operator}
\end{equation}
Considering only the contribution of two-quark operators, the decay width takes the form
\begin{equation}
  \Gamma^{\rm (2q)}_{\rm NL} (B) =
  \left[
    C_1^2 \, \Gamma_{11}^{\rm (2q)}
    + 2 \, C_1 C_2  \,  \left(\frac{1}{N_c} \Gamma_{11}^{\rm (2q)} + 2 \,  \Gamma_{1 T}^{\rm (2q)}  \right) 
   + C_2^2\,  \Gamma_{22}^{\rm (2q)} \right],
   \label{eq:Gamma-B-2q}
\end{equation}
where
\begin{eqnarray}
\Gamma_{11 \boldsymbol{ ( } 1T \boldsymbol{ )} }^{ (2q)} =  
- \frac{4 G^2_F |V_{q_1 b}|^2 |V_{q_2  q_3}|^2}{m_{B}} \,
{\rm Im} \, 
\langle B (p_B)| \,  i \! \int \! d^4 x \, \bar b(0) \, \Gamma_\mu 
\boldsymbol{ ( } t^a \boldsymbol{ )} \, i S^{(q_1)}(0,x)  \Gamma_\nu \, b(x) 
\nonumber \\
\times \,  {\rm Tr} \left[\, \Gamma^\mu\,  \boldsymbol{ ( } t^a \boldsymbol{ ) }\, i S^{( q_3)} (0,x) 
\, \Gamma^\nu \, i S^{(q_2)}(x,0)\, \right] 
|B  (p_B) \rangle  + (x \leftrightarrow 0).
\label{eq:T11}
\end{eqnarray}
Note that in the case of $\Gamma_{ 1T}^{\rm (2q)}$ the two colour matrices $t^a$ appear on the r.h.s of Eq.~(\ref{eq:T11}).
The corresponding expression for $\Gamma_{22}^{\rm (2q)}$ is obtained from that of $\Gamma_{11}^{\rm (2q)}$ 
by replacing $q_1 \leftrightarrow q_3$ as it follows by Fierz-transforming 
the operator $Q_2$ given in Eq.~(\ref{eq.O2-operator}).
To compute dimension-six contributions we need to expand first each of the quark propagators up to one covariant derivative of
the gluon field strength tensor, respectively defined as $i D_\mu = i \partial_\mu + A_\mu(x)$\footnote{Note that the coupling
  constant $g_s$ is absorbed in the definition of $A_\mu = A_\mu^a t^a$.} and  $G_{\mu \nu} = -i \,[i D_\mu, i D_\nu]$.
In Appendix \ref{sec:propagator} we derive the Fourier transform of the quark propagator in the soft external gluon field
\begin{equation}
S(x,0) = \int \frac{d^4 k}{(2 \pi)^4} \, e^{- i k x} \,  S(k),
\label{eq:quark-propagator}
\end{equation}
where
\begin{eqnarray}
S(k) & = & 
\frac{\slashed{k} + m}{k^2-m^2} + 
\frac{1}{2 (k^2-m^2)^2} \left( - m  G_{\rho\mu} \sigma^{\rho\mu} +
2 \, \tilde G_{\rho\mu} \, k^\rho \gamma^\mu \gamma^5 \right) 
\nonumber \\
& & 
+ \, \frac{2}{3} \frac{D_\rho G^{\rho\mu}}{(k^2-m^2)^2} 
  \left( \gamma_\mu - \frac{(\slashed{k} + 2 m) \, k_\mu}{k^2-m^2}\right) 
- \frac{2}{3} \frac{D_\nu G_{\rho\mu} \, k^\nu k^\rho \gamma^\mu}{(k^2-m^2)^3}  
\nonumber\\
& & 
+ \, \frac{2}{3} \frac{m \, D_\nu G_{\rho\mu}}{(k^2-m^2)^3} 
\left( \gamma^\rho \gamma^\mu k^\nu - \gamma^\mu \gamma^\nu k^\rho \right)  
+ 2 i \, \frac{D_\nu \tilde G_{\rho \mu} \, k^\nu k^\rho \gamma^\mu \gamma^5}{(k^2-m^2)^3} + \ldots.
\label{eq:quark-momentum-propagator}
\end{eqnarray}
In the previous equation, $\tilde G_{\rho\mu}= (1/2) \epsilon_{\rho \mu \sigma \eta}G^{\sigma \eta}$ denotes the dual field
strength tensor, with $\epsilon_{\rho\mu\sigma\eta}$ being the Levi-Civita tensor
and $D_\rho G_{\mu\nu} = - [i D_\rho, [i D_\mu, i D_\nu]]$. 
From Eq.~(\ref{eq:quark-momentum-propagator}) it follows that the quark propagator can be split up according to its colour structure as 
\begin{equation}
S_{ij}(k) = S^{(0)} (k) \, \delta_{ij} + S^{(1) \, a} (k) \,  t^a_{ij},
\label{eq:prop-colour-struct}
\end{equation}
where $S^{(0)} (k)$ denotes the free quark propagator and $S^{(1) \, a} (k)$ 
absorbs contributions with one gluon field (for more details see Appendix \ref{sec:propagator}, also the original Refs.~\cite{Blok:1992hw, Blok:1992he}).
This representation allows a straightforward treatment of colour in Eq.~(\ref{eq:T11}). 
In case of $Q_1 \otimes Q_1$ contribution, the colour flow factorises between 
the $(b q_1)$-quark line and the $(q_2 q_3)$-loop, therefore one can only expand the propagator of $q_1$ up to terms linear in $t^a$, see l.h.s. of Fig.~\ref{fig:O1-O1}. 
For the $Q_2 \otimes Q_2$ combination 
the trace over colour indices involves the $q_1$- and $q_2$-quark propagators and 
only the gluon radiation from $q_3$ is non-vanishing, see r.h.s. of Fig.~\ref{fig:O1-O1}.
Finally, in the case of the $Q_1 \otimes T$ contribution, 
the colour flow forces the gluon to be radiated off the $( q_2 q_3)$-loop. 
Any interference term would be ${\cal O}(1/m_b^4)$, and it is thus sufficient to expand independently 
each of the two quark propagators in the loop (Fig.~\ref{fig:O1-O2t}).
\\
Using translation invariance of the quark field to write $b(x) = e^{-ipx}\, b(0)$,
with $p^\mu$ being the $b$-quark four-momentum, and
after performing $x$- and momentum integration, we obtain two-loop tensor integrals of possible rank $r = 1, \ldots , 4$. These are decomposed in terms of rank~$r$ tensors built with the metric tensor $g^{\mu \nu}$ and the external momentum $p^\mu$. The corresponding coefficients represent scalar integrals of the type
\begin{equation} 
\int \! \frac{d^4 k_1}{(2 \pi)^4} \int \! \! \frac{d^4 k_2}{(2 \pi)^4}  
\frac{f (p, k_1, k_2)}{\left[ k_1^2 - m_1^2 \right] ^{n_1} \left[ k_2^2 - m_2^2 \right]^{n_2} \left[(p - k_1 - k_2)^2 - m_3^2 \right]^{n_3}} \,,
\label{eq:two-loop-int}
\end{equation}
where $f (p, k_1, k_2)$ is a function of all the possible scalar products. 
The integrals in Eq.~(\ref{eq:two-loop-int}) are reduced by means of the integration-by-parts (IBP) technique 
to a linear combination of master integrals. 
The IBP reduction is implemented using the Mathematica package 
{\it LiteRed} \cite{Lee:2012cn, Lee:2013mka}.
The discontinuity of the master integrals can be straightforwardly obtained from the imaginary part of 
the scalar sunset diagram $S(s; m_1, m_2, m_3)$, defined by Eq.~(\ref{eq:two-loop-int}) 
with $f = 1$ and $n_1 = n_2 = n_3 = 1$ and given by
\footnote{For instance, the discontinuity of the master integral with $f = 1$, $n_1 = 2$ and $n_2 =n_3=1$ can be obtained
by differentiating Eq.~(\ref{eq:Disc-scalar-integral}) with respect to $m_1^2$, etc.}, see Ref.~\cite{Remiddi:2016gno}
\begin{equation}
{\rm Im} \, S (s; m_1, m_2, m_3) = \frac{1}{256 \pi^3}
\! \! \! \! \int\limits_{(m_2 + m_3)^2}^{(\sqrt{s} - m_1)^2} 
\! \! d t \, \frac{\sqrt{\lambda(t, m_2^2, m_3^2)\, \lambda(s, t, m_1^2)}}{t \, s}.
\label{eq:Disc-scalar-integral}
\end{equation}
In the previous equation, $s = p^2 \geq (m_1 + m_2 + m_3)^2$ and $\lambda(a,b,c) \equiv a^2 + b^2 + c^2 - 2 (ab + bc + ac)$ is the K\"allen function.
One can easily compute analytically the integral in Eq.~(\ref{eq:Disc-scalar-integral}) for two non-vanishing masses, while
in the case of three non-vanishing masses the complexity highly increases and the solution involves elliptic functions, see e.g. Refs.~\cite{Pivovarov:1984ij,Groote:2000kz,Broedel:2017kkb,Broedel:2017siw}. 
We emphasize that we set $d = 4$ from the beginning since the discontinuity of the diagrams 
at LO-QCD does not develop any ultraviolet divergences.
On the other side, the gluon emission off a light quark propagator ($q =u ,d$ or $s$) 
gives rise, at dimension-six, to infrared logarithmic divergences of the type $\log (m_{q}^2/m_b^2)$ 
which are cancelled by the renormalised one-loop matrix element of the corresponding  four-quark operators,
as shown in detail in Section \ref{sec:logs}.
Here we limit ourselves to state that for the computation of most of the diagrams of interest it is always
possible to set one mass to zero and obtain an analytic expression for all the master integrals. 
This is not the case only for the gluon emission from the $s$ quark in the $b \to c \bar c s$ transition, 
where we need to keep all masses finite in order to regularise the infrared divergence.
\\
After computing the two-loop integrals we are left with the following matrix elements 
\begin{eqnarray}
&& \langle B (p_B)| \bar b_v (0) \, {\cal F}(p )\,  b_v (0)| B (p_B) \rangle,
\label{ME1} \\
&&\langle B (p_B)| \bar b_v (0)\, {\cal F}_{\mu \nu}(p) \,  
(i D^\mu) (i D^\nu) b_v (0)| B (p_B) \rangle,
\label{ME2} \\
&&\langle B (p_B)| \bar b_v (0)\, {\cal F}_{\mu \nu \rho}(p) \,  
(i D^\mu) (i D^\nu) (i D^\rho) b_v (0)|B (p_B) \rangle, 
\label{ME3}
\end{eqnarray}
where ${\cal F} (p)$, ${\cal F}_{\mu \nu} (p)$ and ${\cal F}_{\mu \nu \rho} (p)$ 
are functions of the quark masses and of the external momentum
$p^\mu$.
 We find that 
 at order $1/m_b^2$ only the expansion of the anti-quark propagator contributes
 (in agreement with Refs.~\cite{Bigi:1992su,Blok:1992hw,Blok:1992he,Bigi:1992ne})
 and therefore the matrix element of dimension-five, Eq.~(\ref{ME2}), vanishes in
 the case of both $Q_1 \otimes Q_1$ and $Q_2 \otimes Q_2$ contributions but not for $Q_1 \otimes Q_2$. 
On the other hand, the matrix elements of dimension-three,
 Eq.~(\ref{ME1}), and of dimension-six, Eq.~(\ref{ME3}), are non-vanishing for all three combinations of $\Delta B = 1$ operators.
  
\noindent  
Following the standard technique of the HQE we decompose
the $b$-quark momentum~as
\begin{equation}
p^\mu = m_b v^\mu + i D^\mu,
\label{eq.HQE-momentum}
\end{equation} 
where $v^\mu$ represents the four-velocity of the $B$ meson and $D^{\mu}$ accounts for the soft interaction with the spectator quark. 
We have used the phase redefinition
\begin{equation}
b(x) = e^{-i m_b v \cdot x} b_v(x),
\label{eq.bv-definition}
\end{equation}
to remove the large fraction of the $b$-field momentum, at $x=0$ we trivially get $b(0) = b_v(0)$.
We then expand Eqs.~(\ref{ME1})-(\ref{ME3}) in the small quantity
$D^\mu/m_b$.
Note that in the matrix elements with three covariant derivatives we can safely set $p^2 = m_b^2$
neglecting corrections ${\cal O} (1/m_b^4)$. 
The order of the covariant derivatives is fixed by 
\begin{equation}
p^{\mu_1} p^{\mu_2} \ldots p^{\mu_n} = 
\frac{1}{n!} \sum_{\sigma \in {\cal S}_n} p^{\sigma(\mu_1)} p^{\sigma(\mu_2)} \ldots p^{\sigma(\mu_n)},
\end{equation} 
where ${\cal S}_n$ is the symmetric group of permutation of $n$ elements.
This leads to a systematic expansion of Eq.~(\ref{eq:T11}), schematically 
\begin{equation}
  a \langle \bar b_v b_v \rangle + 
  b_\mu \langle \bar b_v (i D^\mu) b_v  \rangle + 
  c_{\mu \nu} \langle \bar b_v (i D^\mu) (i D^\nu) b_v \rangle + 
  d_{\mu \nu \rho} \langle \bar b_v (i D^\mu) (i D^\nu) (i D^\rho) b_v  \rangle + \ldots, 
\label{eq.matrix-element-decomposition}
\end{equation}
where $a, b_\mu, c_{\mu\nu}, d_{\mu\nu\rho}$ are now only functions of the quark masses 
and of the four-velocity $v^\mu$. \\
Finally, the matrix elements in Eq.~(\ref{eq.matrix-element-decomposition}) 
admit a series expansion 
in powers of $1/m_b$ given explicitly in Ref.~\cite{Dassinger:2006md}\footnote{For reference we quote the expansion
of the dimension-three matrix element up to order $1/m_b^3$, $\langle  B (p_B)| \bar b_v  b_v | B (p_B) \rangle = 2 m_{B} \left(1- \frac{\mu_\pi^2(B)- \mu_G^2(B)}{2 m_b^2} \right)$.},
from which we can readily 
obtain the coefficients of $\mu_\pi^2, \mu_G^2$ and of $\rho_D^3, \rho_{LS}^3$, which are defined as
\begin{eqnarray}
2 m_B \, \mu_\pi^2 (B) & = & 
- \langle B (p_B) | \bar b_v (i D_\mu) (i D^\mu) b_v | B (p_B) \rangle,
\nonumber\\
2 m_B \, \mu_G^2 (B) & = & 
\langle B (p_B) | \bar b_v (i D_\mu)  (i D_\nu) 
(-i \sigma^{\mu\nu}) b_v | B (p_B) \rangle, 
\label{eq:NP-par-dim-5}
\\[1mm]
2 m_B \, \rho_D^3 (B)  & = & 
\langle B (p_B) | \bar b_v (i D_\mu) (i v \cdot D) (i D^\mu)b_v | B (p_B) \rangle, \nonumber\\
2 m_B \, \rho_{LS}^3 (B)  & = & 
\langle B (p_B) | \bar b_v (i D_\mu) (i v \cdot D) (i D_\nu) (-i \sigma^{\mu\nu})\, b_v | B (p_B) \rangle,
\label{eq:NP-par-dim-6}
\end{eqnarray}
with $\sigma_{\mu \nu} = (i/2) [\gamma_\mu, \gamma_\nu]$.
As already stressed, the numerical values of the non-perturbative parameters above depend on the spectator quark in the $B$
meson.
\\
We then arrive at the following form of Eq.~(\ref{eq:Gamma-B-2q})
\begin{eqnarray}
\Gamma^{\rm (2q)}_{\rm NL} (B) & = & \Gamma_0 
\biggl[\left(3 \, C_1^2  + 2 \, C_1  C_2  + 3 \, C_2^2 \right) {\cal C}_0^{(q_1 \bar q_2 q_3)} 
\left(1 - \frac{\mu_\pi^2 (B)} {2 m_b^2} \right) 
\nonumber \\
&  & + \left(3 \, C_1^2 \,\, {\cal C}_{G, 11}^{(q_1 \bar q_2 q_3)} 
+ 2 \, C_1  C_2 \,\, {\cal C}_{G, 12}^{(q_1 \bar q_2 q_3)} 
+ 3 \, C_2^2 \,\, {\cal C}_{G, 22}^{(q_1 \bar q_2 q_3) } \right) \frac{\mu_G^2 (B)} {m_b^2}  
\nonumber \\
& & +\left(3 \, C_1^2 \,\, {\cal C}_{D, 11}^{(q_1 \bar q_2 q_3)} 
+ 2 \, C_1  C_2 \,\, {\cal C}_{D,12}^{(q_1 \bar q_2 q_3)} 
+ 3 \, C_2^2 \,\, {\cal C}_{D, 22}^{(q_1 \bar q_2 q_3) } \right) \frac{\rho_D^3 (B)} {m_b^3}  
\biggr],
\label{eq:Gamma-NL-res-scheme}
\end{eqnarray}
where
\begin{equation}
\Gamma_0 = \frac{G_F^2 m_b^5}{192 \, \pi^3} |V_{q_1 b}|^2 |V_{q_2  q_3}|^2.
\label{eq:Gamma0}
\end{equation}
In Eq.~(\ref{eq:Gamma-NL-res-scheme}), ${\cal C}_0^{(q_1 \bar q_2 q_3)}$ refers to the partonic-level
coefficient, while ${\cal C}_{G, nm}^{(q_1 \bar q_2 q_3)}$ and ${\cal C}_{D,  nm}^{(q_1 \bar q_2 q_3)}$,  
$nm = \{11, 12, 22\}$, denote the coefficients of the chromo-magnetic and of the Darwin operators,
respectively.
The upper index $(q_1 \bar q_2 q_3)$ indicates the decay mode, e.g. $(c \bar u d)$ 
corresponds to the $b \to c \bar u d$ transition.
Since we neglect the masses of the up, down and strange quark, all the coefficients in Eq.~(\ref{eq:Gamma-NL-res-scheme}) are functions of at most one
dimensionless mass parameter
 $\rho = m_c^2/m_b^2$, apart from ${\cal C}_{D,mn}^{(q_1 \bar q_2 q_3)}$, where the dependence on the light quark mass $m_{q} = m_{u,d,s}$
  is still present in the form of divergent logarithms $ \log (m_q^2/m_b^2)$. The cancellation of these terms is discussed in the next subsection.
The complete LO-QCD expressions for ${\cal C}_0^{(q_1 \bar q_2 q_3)}$ and ${\cal C}_{G, nm}^{(q_1 \bar q_2 q_3)}$ can be found in Appendix \ref{App:C3C5}.
We point out that the adopted definition for the non-perturbative parameters in
Eqs.~(\ref{eq:NP-par-dim-5}), (\ref{eq:NP-par-dim-6}) implies that 
the coefficient of $\rho_{LS}^3$ is found to vanish for all the $\Delta B = 1$ 
operator combinations.

\subsection{Role of the four-quark operators}
\label{sec:logs}
\begin{figure}[t]\centering
\includegraphics[scale=0.45]{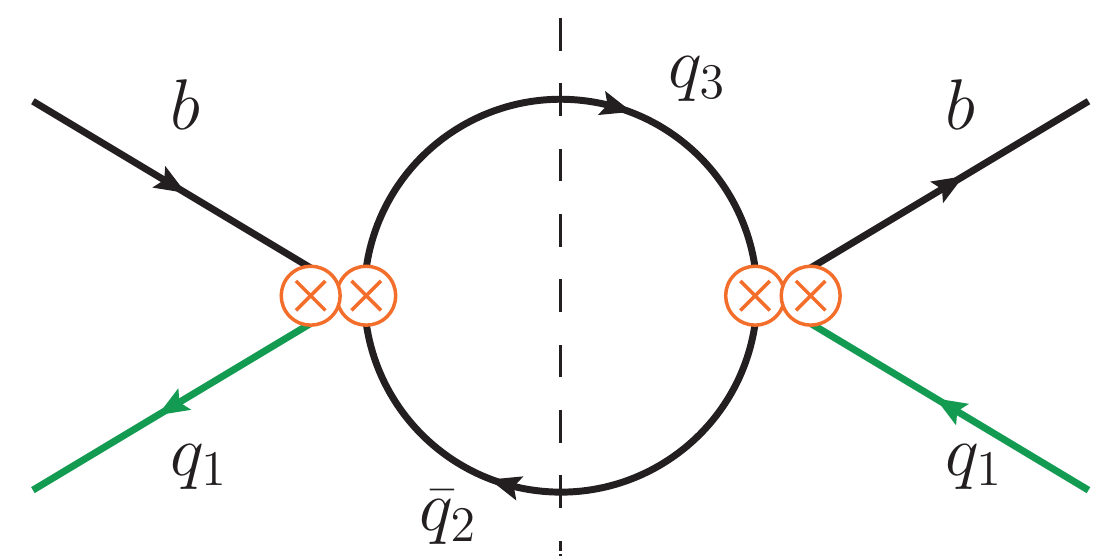}
\qquad
\includegraphics[scale=0.45]{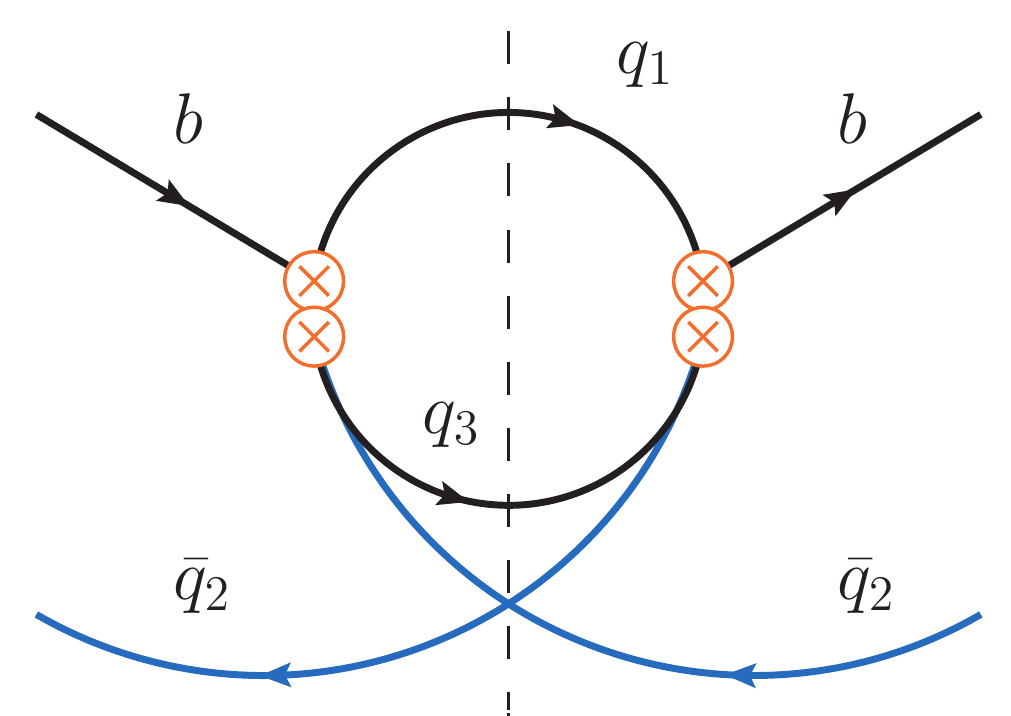}
\qquad
\includegraphics[scale=0.45]{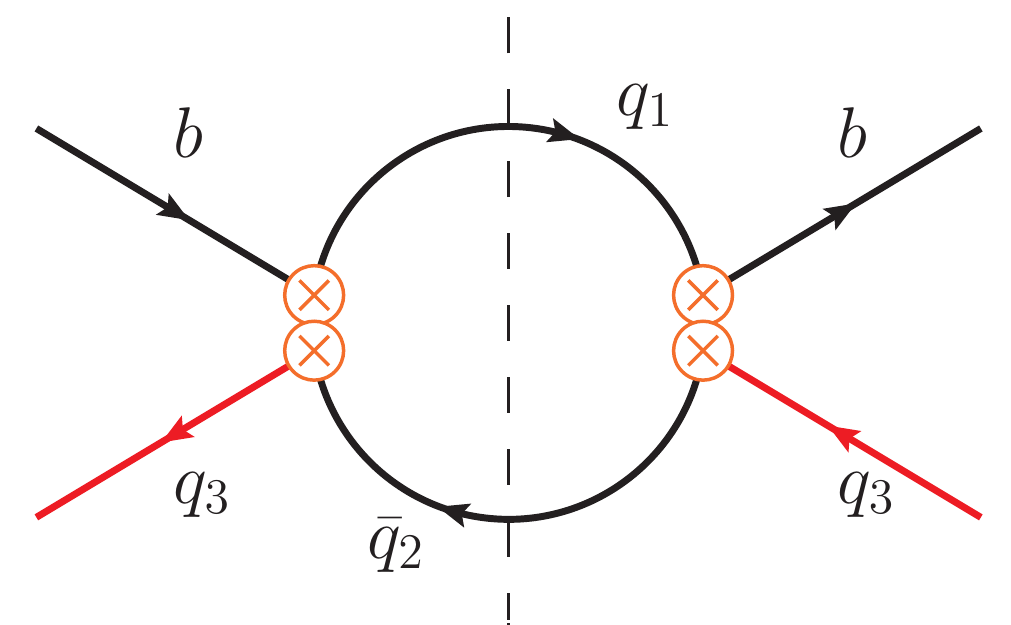}
\caption{One-loop diagrams corresponding, from left to right, to the WA, PI and WE~topology.}
\label{fig:PI-and-WE}
\end{figure}
At order $1/m_b^3$ the gluon radiation from a light quark leads to IR divergences, namely
\begin{equation}
{\cal C}_{D,nm}^{(q_1 \bar q_2 q_3)} 
=  {\cal R}_{nm}^{(q_1 \bar q_2 q_3)} + {\cal D}_{nm}^{(q_1 \bar q_2 q_3)} 
\left( \log \left(\frac{m_{q}^2}{m_b^2} \right) \right),
\label{eq:C-rho-D-lim}
\end{equation}
where ${\cal R}_{nm}^{(q_1 \bar q_2 q_3)}$ are finite functions and 
${\cal D}_{nm}^{(q_1 \bar q_2 q_3)} \left(\log (m_{q}^2/m_b^2) \right)$ 
absorb the contribution of the divergent logarithms, 
the latter are listed in Appendix \ref{App:Mixing}.
As discussed already in Ref.~\cite{Novikov:1983gd},
logarithmic infrared divergences signal mixing between operators of the same dimension. 
To~see this in more detail, we start again from Eq.~(\ref{eq:Gamma-def}),
\footnote{For brevity we omit the indices $(q_1 \bar q_2 q_3)$ and $(n m)$, although they must be always understood.} 
\begin{equation}
\Gamma_{\rm NL} (B) = 
\frac{1}{2 m_{B}} \langle B(p_B) |\,  {\rm Im} \, \hat T\,  | B (p_B) \rangle,
\end{equation}
with the transition operator $\hat T$ being the time-ordered product of the double insertion 
of the effective Lagrangian.
\begin{figure}[t]\centering
\includegraphics[scale=0.63]{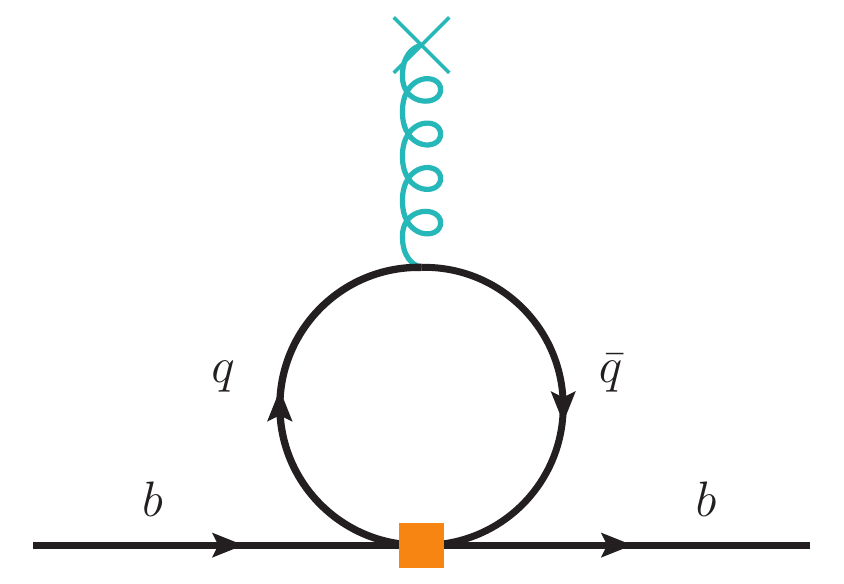}
\caption{Diagram contributing to mixing between four- and two-quark operators at dimension-~six.}
\label{fig:ME-O-T-1-2}
\end{figure}
\noindent
The expansion in inverse powers of $m_b$ allows 
to express $\hat T$ in terms of new local effective operators. 
At dimension-six one has \footnote{In principle the sum in Eq.~(\ref{eq:OPE}) includes also the $c$-quark but 
for $m_b \sim m_c \gg \Lambda_{QCD}$ the four-quark operators~$ \tilde{\cal O}^{(c)}_{6,i} $ 
are not relevant for the further discussion, see for more details Ref.~\cite{Breidenbach:2008ua}.}: 
\begin{equation}
{\rm Im} \,  \hat T^{(d = 6)}  
= \Gamma_0 \,  C (\mu_1) 
\left[ 
{\cal C}_{\rho_D} (\mu_1, \mu_0)  \frac{{\cal O}_{\rho_D} (\mu_0)}{m_b^3} 
+ \sum_{q= u,d,s} \sum_{i = 1}^{4} {\cal C}^{(q)}_{6,i} (\mu_1, \mu_0) 
\frac{\tilde{\cal O}^{(q)}_{6,i} (\mu_0)}{m_b^3} 
\right],
\label{eq:OPE}
\end{equation}
where $C(\mu_1) = \{ 3\, C_1^2, 2 \,C_1 \, C_2, 3 \, C_2^2\}$ and $\mu_1$ is the renormalisation scale at which the $\Delta B = 1$ Wilson coefficients
of the weak Hamiltonian are determined. 
This scale arises also in loop corrections to the diagrams given in the upper line of Fig.~\ref{fig:intro}.
Up to the calculated order in the strong coupling the $\mu_1$  dependence will cancel among these two sources.
The second scale~$\mu_0$ ($\Lambda_{\rm QCD} \ll \mu_0 \le \mu_1 \sim m_b $) indicates the new factorisation scale of
the $\Delta B = 0 $ operators
emerging in the HQE. 
The $\mu_0$ dependence will cancel among loop corrections to the diagrams given in the lower line of Fig.~\ref{fig:intro} and the
scale dependence of the $\Delta B = 0$ operators.
\\
The l.h.s of Eq.~(\ref{eq:OPE}) includes the contributions given by the two-loop diagrams 
in Fig.~\ref{fig:O1-O1} and Fig.~\ref{fig:O1-O2t}, 
which give rise directly to the Darwin operator ${\cal O}_{\rho_D}$,
\footnote{We do not take into account the spin-orbit operator
$
{\cal O}_{\rho_{LS}} = \bar b_v (i D_\mu) (i v \cdot D) (i D_\nu) 
(-i \sigma^{\mu\nu})\, b_v
$
since its contribution is vanishing within the adopted convention, see Eq.~(\ref{eq:NP-par-dim-6}). }
\begin{eqnarray}
{\cal O}_{\rho_D} & = & 
\bar b_v (i D_\mu) (i v \cdot D) (i D^\mu)b_v \,,
\label{eq:2q-operators}
\end{eqnarray}
with the corresponding coefficient ${\cal C}_{D} (\mu_1)$ - see  Eq.~(\ref{eq:Gamma-NL-res-scheme}).
Since we are only working at LO-QCD, we do not find
any explicit $\mu_1$-dependence at this stage. The l.h.s of Eq.~(\ref{eq:OPE}) receives also contributions
by the one-loop diagrams depicted in Fig.~\ref{fig:PI-and-WE},
corresponding to the weak-annihilation, Pauli interference
and weak-exchange topologies. 
The  coefficients ${\cal C}^{(q)}_{6,i} (\mu_1, \mu_0)$ 
do not develop any divergences at LO-QCD
\cite{Uraltsev:1996ta, Neubert:1996we} 
hence there is no explicit $\mu_1$- and $\mu_0$-dependence; 
it is however present in the NLO-QCD corrections determined in Refs.~\cite{Ciuchini:2001vx,Beneke:2002rj,Franco:2002fc}.
We refer to Appendix~\ref{App:Mixing} for  explicit expressions at LO-QCD.
Integrating out these one-loop diagrams leads to the following 
$\Delta B = 0$ four-quark operators:
\footnote{
In the literature typically colour singlet and colour octet operators are used,
see e.g. Refs.~\cite{Uraltsev:1996ta,Neubert:1996we,Lenz:2015dra,Kirk:2017juj}; 
however, for our purposes it turns out to be advantageous to use instead 
the colour singlet and the colour rearranged operators. 
The trivial transformation between the two bases is given by Eq.~(\ref{eq:colour-trafo}).
}
\begin{eqnarray}
\hspace*{-5mm} \tilde{\cal O}_{6,1}^{(q)} & = & 
(\bar b_v^i \gamma_\mu (1 - \gamma_5) \, q^i )
(\bar q^{\, j} \gamma^\mu (1 - \gamma_5)  \, b_v^{\, j}), 
\quad
\tilde {\cal O}_{6,2}^{(q)}  = 
(\bar b_v^i \, \slashed v (1 - \gamma_5)  \, q^i)(\bar q^{\, j} \, \slashed v (1 - \gamma_5)  \,b_v^{\, j}), 
\quad
\label{eq:4q-operators-O12} \\
\hspace*{-5mm}
\tilde{\cal O}_{6, 3}^{(q)} & = & 
( \bar b_v^i \gamma_\mu (1 - \gamma_5)  \, q^{\, j} )(\bar q^j \gamma^\mu (1 - \gamma_5) \, b_v^i),  \quad 
\tilde{\cal O}_{6, 4}^{(q)}  = 
(\bar b_v^i\, \slashed v (1 - \gamma_5)  \, q^{\, j})(\bar q^{\, j} \, \slashed v (1 - \gamma_5)  \,b_v^i).
\quad
\label{eq:4q-operators-O34}
\end{eqnarray}
\noindent
The four-quark and the Darwin operators mix under operator renormalisation already at LO-QCD, through the diagram in Fig.~\ref{fig:ME-O-T-1-2}. In dimensional regularisation with $d = 4 - 2 \epsilon$ a factor of $\mu^{2 \epsilon}$ must be included in order to compensate the mismatch in the dimension of $\tilde{\cal O}_{6,i}^{(q)}$ and ${\cal O}_{\rho_D}$, leading to explicitly $\mu$-dependent renormalisation constants and thus a non-vanishing anomalous dimension matrix.
The renormalised one-loop matrix element of $\tilde{\cal O}_{6,i}^{(q)}$ in the presence of a soft background gluon field,
in NDR and in the $\overline{ {\rm MS}}$ scheme reads:
\begin{eqnarray}
\langle \,  \tilde{\cal O}_{6,i}^{(q)} \, \rangle^{\rm ren} & = &  
\frac{a_i}{12 \pi^2} \left[\log \left( \frac{\mu^2}{m_q^2} \right) + b_i \right]
     \langle {\cal O}_{\rho_D} \rangle 
+ {\cal O} \left( \frac{1}{m_b} \right)
\label{eq:ME-OT1} ,
\end{eqnarray}
with $a_1 = 2$, $a_2 = -1$, $a_3 = a_4 = 0$ and $b_1 = -1$ and $b_2 = 0$.
Note that the presence of the constant term $b_i$ depends 
on the choice of the operator basis in Eqs.~(\ref{eq:4q-operators-O12}), 
(\ref{eq:4q-operators-O34}), see for instance Refs.~\cite{Bigi:2005bh, Gambino:2005tp}.
At the matching scale $\mu  = m_b $, we obtain
\begin{equation}
    {\cal C}_{\rho_D}(\mu_1, m_b) \, \langle \, {\cal O}_{\rho_D}(m_b) \, \rangle 
    = {\cal C}_{D}    (\mu_1) \, \langle \, {\cal O}_{\rho_D}(m_b) \, \rangle
     - \sum_{q = u,d,s} \sum_{i=1}^2 \,
     {\cal C}_{6,i}^{(q)}  (\mu_1, m_b) \,  \langle \,
  \tilde{\cal O}_{6,i}^{(q)}(m_b) \, \rangle^{\rm ren},
\label{eq:matching-mb}
\end{equation}
where the coefficient ${\cal C}_{D}$ has the divergent
logarithmic dependence shown in Eq.~(\ref{eq:C-rho-D-lim}). From 
Eq.~(\ref{eq:ME-OT1}),
it follows that on the r.h.s of Eq.~(\ref{eq:matching-mb}) all the logarithms $\log(m_{q}^2/m_b^2)$ cancel exactly,
leaving ${\cal C}_{\rho_D}(\mu_1, m_b)$ free of any IR divergences. 
Finally one
can solve the renormalisation group equations to run ${\cal C}_{\rho_D}(\mu_1, m_b)$
down to the scale $m_b \geq \mu_0 \gg \Lambda_{\rm QCD}$:
\begin{equation}
  {\cal C}_{\rho_D}(\mu_1, \mu_0) = {\cal C}_{\rho_D}( \mu_1, m_b) - \frac{1}{12 \pi^2} \log \left( \frac{\mu_0^2}{m_b^2} \right ) \sum_{q=u,d,s}
  \left[2 \,   {\cal C}_{6,1}^{(q)}( \mu_1, m_b) -  \, {\cal C}_{6,2}^{(q)}(\mu_1,  m_b) \right].
\label{eq:C-rho-D-mu}
\end{equation}
The coefficients ${\cal C}_{\rho_D}(\mu_1, \mu_0)$ are now IR-finite but scale dependent. This $\mu_0$-dependence cancels the corresponding one due to the running of the matrix element of the four-quark operators starting already at LO-QCD via the diagram in Fig.~\ref{fig:ME-O-T-1-2}:
\begin{equation}
 \langle \tilde {\cal O}_{6, i}^{(q)}(\mu_0) \rangle = \langle \tilde {\cal O}_{6, i}^{(q)}(m_b) \rangle + \frac{a_i}{12 \pi^2} \log \left(\frac{\mu_0^2}{m_b^2} \right) \langle {\cal O}_{\rho_D}(m_b) \rangle.
\end{equation}
Our results are presented in the next section.
We conclude with the remaining dimension-six contribution to $\Gamma_{\rm NL}(B)$
(see Appendix~\ref{App:Xs} for the expression of the coefficients at LO-QCD):
\begin{equation}
  \Gamma_{\rm NL}^{( \rm 4 q)} (B) =
  \frac{\Gamma_0}{2 m_B} \, C(\mu_1) \sum_{q = u,d,s} \sum_{i = 1}^{4} {\cal C}^{(q)}_{6,i}(\mu_1, \mu_0) \,
  \frac{ \langle \tilde{\cal O}^{(q)}_{6,i}  (\mu_0) \rangle}{m_b^3},
\end{equation}
where the matrix element of the four-quark operators can be parametrised as
\begin{eqnarray}
\langle B_{q^\prime} |\, \tilde{\cal O}_{6,i}^{(q)} \, |B_{q^\prime} \rangle  =  A_i \, m_{B}^2 \,  f_B^2 \left(
          {\cal B}_i^{(q)}(B) \,  \delta_{q q^\prime} + 
          \tau_i^{(q)}(B)
           \right), \qquad q^\prime = u,d,s,
\label{eq:NP-parameters-B-t}
\end{eqnarray}
with $A_1=A_3 = 1$, $A_2 = A_4 = (m_{B}/(m_b + m_{q}))^2$ 
and $f_B$ being the decay constant of the $B$~meson.
We have separated the contribution due to the valence and non-valence quarks, 
where ${\cal B}_i^{(q)}$ is non vanishing only for $q$ equal to $q^\prime$, 
the spectator quark in the $B$ meson, and is expected to be of order one, 
see e.g. Ref.~\cite{Kirk:2017juj}, while $\tau_i^{(q)}$ accounts for the effects of 
an "intrinsic" $q$~quark \cite{Breidenbach:2008ua, Fael:2019umf} 
and is expected to be small. 
Its numerical value can be estimated via e.g. the calculation of the so-called eye contractions 
in the non-perturbative determination of the matrix elements, see Ref. \cite{King:2020}.

\section{Results}
\label{sec:results}
\begin{figure}[t]
\includegraphics[scale=0.7]{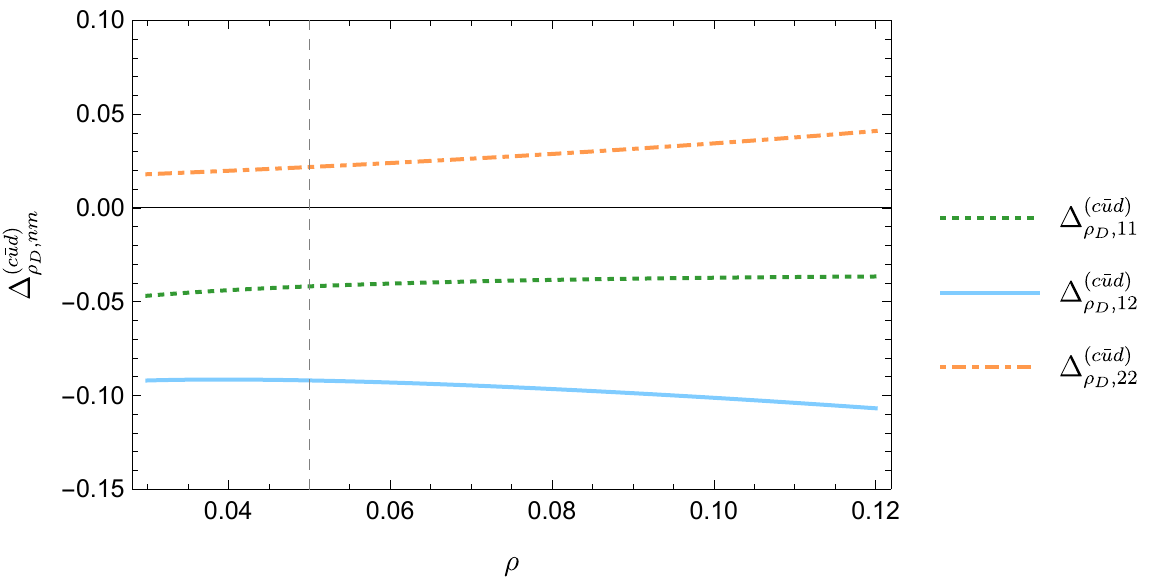} \quad
\includegraphics[scale=0.7]{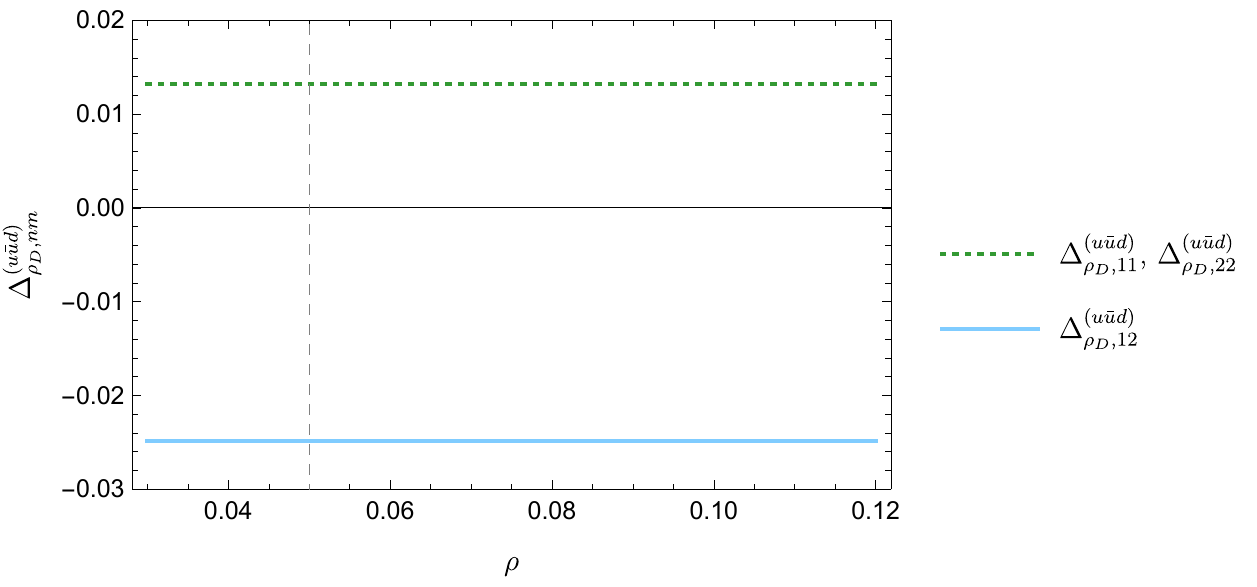}

\includegraphics[scale=0.7]{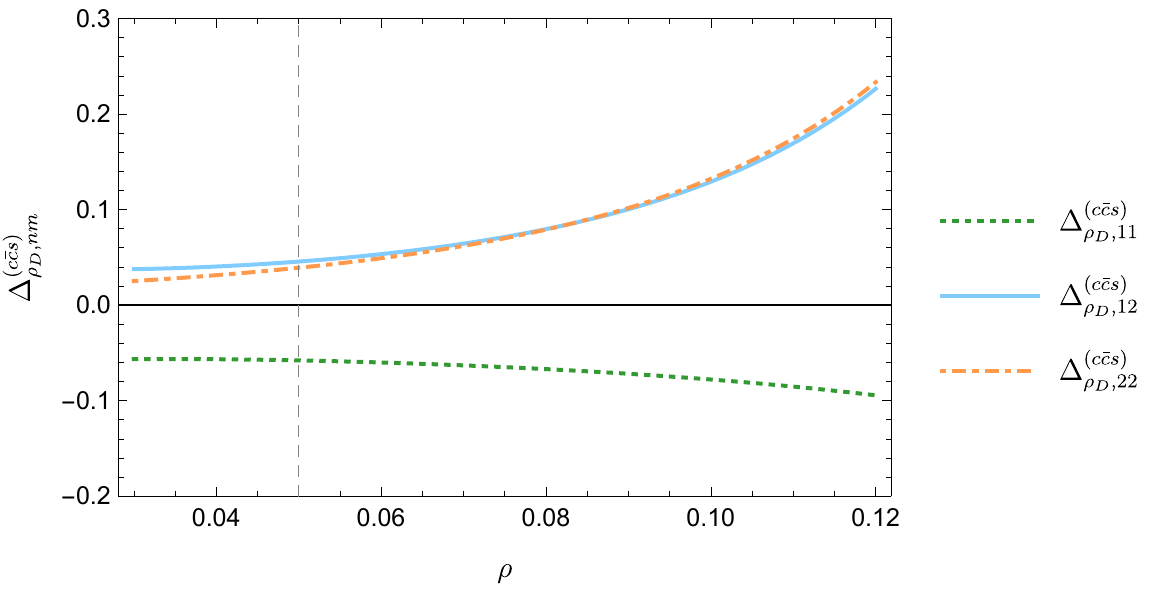} \quad
\includegraphics[scale=0.7]{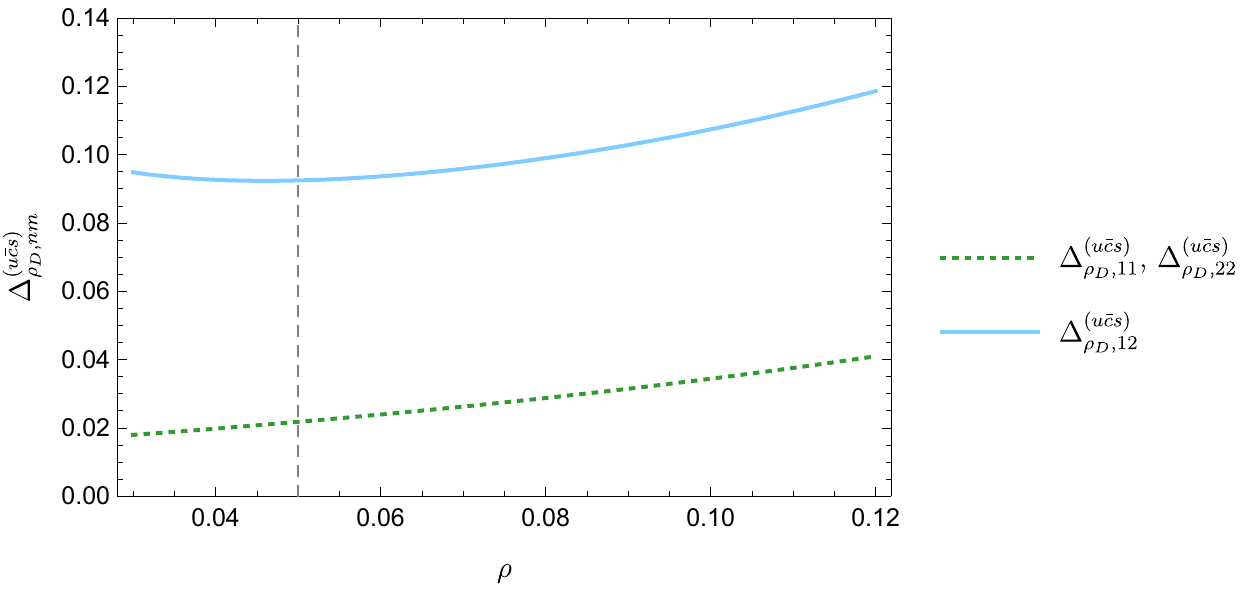}
\caption{The Darwin term correction normalised to the partonic-level coefficient for different modes, namely $b \to c \bar u d$ (top left),
  $b \to u \bar u d$ (top right), $b \to c \bar c s$ (bottom left), and $b \to u \bar c s$ (bottom right).
  The green dotted line corresponds to $Q_1 \otimes Q_1$, the solid cyan line to $Q_1 \otimes Q_2$ and
  the dotted-dashed orange line to the $Q_2 \otimes Q_2$ contribution. For reference we have fixed the
  values $\mu_0 = m_b$, $m_b=4.5$ GeV and $\rho_D^3 = 0.2$ GeV$^3$. The dashed vertical line shows the approximate
  value $\rho = 0.05$ in the $\overline{ {\rm MS}}$ scheme.
}
\label{fig:Deltas}
\end{figure}

\begin{figure}[t]\centering
\includegraphics[scale=1.0]{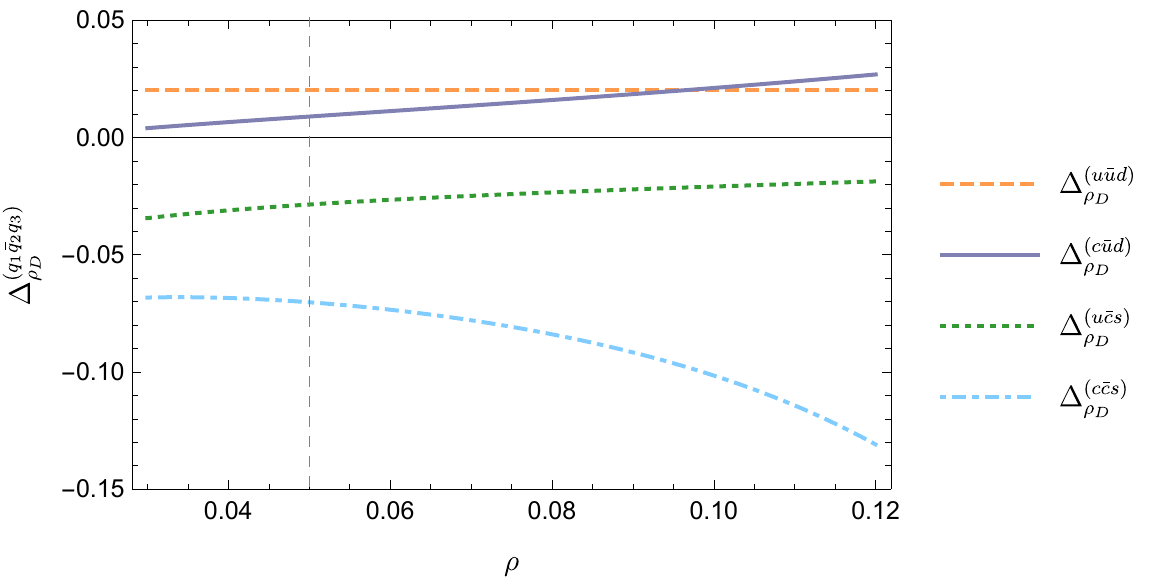} 
\caption{Total relative size of the Darwin term correction compared to the partonic-level contribution for  different modes,
  including  $b \to u \bar u d$ (dashed orange), $b \to c \bar u d$ (solid purple), $b \to u \bar c s$ (dotted green)
  and $b \to c \bar c s$ (dot-dashed cyan). For reference we have fixed the values $\mu_1 = \mu_0 = m_b$, $m_b=4.5$ GeV and $\rho_D^3 = 0.2$~GeV$^3$. 
  The dashed vertical line shows the approximate value $\rho = 0.05$ 
  in the $\overline{ {\rm MS}}$ scheme.
}
\label{fig:Deltas-sum}
\end{figure}

\noindent
The contribution of the Darwin operator to the inclusive non-leptonic decay
$b \to q_1 \bar q_2 q_3$ is presented in the following form
\begin{eqnarray}
\Gamma^{(\rho_D)}_{\rm NL} (B) & = &
\Gamma_0 \, \Bigl( 3 \, C_1^2 \, {\cal C}_{\rho_D, \, 11}^{(q_1 \bar q_2 q_3)} 
+ 2 \, C_1 C_2 \, {\cal C}_{\rho_D,\, 12}^{(q_1 \bar q_2 q_3)} 
+ 3 \, C_2^2  \, {\cal C}_{\rho_D,\, 22}^{(q_1 \bar q_2 q_3) }  \Bigr) \frac{\rho_D^3}{m_b^3},
\label{eq:Gamma-NL-final-res}
\end{eqnarray}
where $\rho_D^3$ is defined in Eq.~(\ref{eq:NP-par-dim-6}),
$C_{1,2} (\mu_1)$ are the $\Delta B = 1 $ Wilson coefficients, and
\begin{eqnarray}
{\cal C}_{\rho_D, 11}^{(u \bar u d)} 
& = &
6 + 8 \, \log \left(\frac{\mu_0^2}{m_b^2}\right),
\\ 
{\cal C}_{\rho_D, 12}^{(u \bar u d)} 
& = & 
-\frac{34}{3}, 
\\
{\cal C}_{\rho_D, 22}^{(u \bar u d)} 
& = &
6 + 8 \, \log \left(\frac{\mu_0^2}{m_b^2}\right),
\label{eq:Ruud}
\end{eqnarray}

\begin{eqnarray}
{\cal C}_{\rho_D, 11}^{(u \bar c s)} 
& = & 
\frac{2}{3} (1 - \rho) \biggl[ 9  + 11 \rho - 12 \rho ^2 \log (\rho ) 
 - \, 24 \left(1 - \rho^2 \right) \log (1-\rho )- 25 \rho ^2  + 5 \rho^3 \biggl]
\nonumber \\
& & 
 + \,  8 \, (1 - \rho)^2 (1 + \rho) \log \left(\frac{\mu_0^2}{m_b^2}\right) , 
\label{eq:R11ucs} 
\\
{\cal C}_{\rho_D, 12}^{(u \bar c s)} 
& = & 
\frac{2}{3} \biggl[ - 41 - 12 \left(2 + 5 \rho + 2 \rho ^2 - 2  \rho ^3 \right) \log( \rho) 
\nonumber \\
& &
\quad \quad  - \, 48 (1 - \rho)^2 (1 + \rho ) \log (1-\rho) + 26  \rho - 18 \rho^2 + 38 \rho^3 - 5 \rho ^4 \biggl]
\nonumber \\
& &
+\, 16 \, (1 - \rho)^2 (1 + \rho) \log \left(\frac{\mu_0^2}{m_b^2}\right), 
\label{eq:R12ucs} 
\\
{\cal C}_{\rho_D, 22}^{(u \bar c s)} 
& = &
\frac{2}{3} (1 - \rho) \biggl[ 9  + 11 \rho - 12 \rho ^2 \log (\rho )  - \, 24 \left(1 - \rho^2 \right) \log (1-\rho )- 25 \rho ^2  + 5 \rho^3 \biggr]
\nonumber \\ 
& & 
+\, 8 \, (1 - \rho)^2 (1 + \rho) \log \left(\frac{\mu_0^2}{m_b^2}\right),
\label{eq:R22ucs}
\end{eqnarray}
\\[5mm]

\begin{eqnarray}
{\cal C}_{\rho_D, 11}^{(c \bar u d)} 
& = & 
\frac{2}{3} \biggl[17 +12 \log (\rho)-16 \rho -12 \rho ^2 + 16 \rho ^3 - 5 \rho ^4 \biggr], 
\label{eq:R11cud} 
\\  
{\cal C}_{\rho_D, 12}^{(c \bar u d)} 
& = & 
\frac{2}{3} \biggl[- 9 + 12 \left(1 - 3 \rho ^2 + \rho ^3 \right) \log( \rho) 
\nonumber \\
& & 
\quad \quad + \, 24 (1 - \rho)^3 \log (1-\rho) +  50 \rho - 90 \rho ^2 + 
54 \rho ^3 - 5 \rho ^4 \biggr] \hspace*{25mm}
\nonumber \\
& & - \, 8 \, (1 - \rho)^3 \log \left(\frac{\mu_0^2}{m_b^2}\right),  
\label{eq:R12cud} 
\\
{\cal C}_{\rho_D, 22}^{(c \bar u d)} 
& = &
\frac{2}{3} (1 - \rho) \biggl[ 9  + 11 \rho - 12 \rho ^2 \log (\rho ) 
\nonumber \\
& & 
\quad \quad - \, 24 \left(1 - \rho^2 \right) \log (1-\rho )- 25 \rho ^2  + 5 \rho^3 \biggr]
\nonumber \\
& & + \, 8 \, (1 - \rho)^2 (1 + \rho) \log \left(\frac{\mu_0^2}{m_b^2}\right) ,
\label{eq:R22cud}
\end{eqnarray}

\begin{eqnarray}
{\cal C}_{\rho_D, 11}^{(c \bar c s)} & = & 
\frac{2}{3} \Biggl[ \sqrt{1 - 4 \rho} \left(17 + 8 \rho - 22 \rho^2 - 60 \rho^3 \right)
\nonumber \\
& & \quad \quad - \, 12 \left(1 - \rho - 2 \rho^2 + 2 \rho^3 + 10 \rho^4 \right) 
\log \left(\frac{1 + \sqrt{1 - 4 \rho^{\phantom{\! 1}}}}
{1 - \sqrt{1 - 4 \rho^{\phantom{\! 1}}}} \right) \Biggr],
 \label{eq:R11ccs} \\
{\cal C}_{\rho_D, 12}^{(c \bar c s)} & = & 
\frac{2}{3} \Biggl[ \sqrt{1 - 4 \rho} \left(-45 + 46 \rho - 106 \rho^2 - 60 \rho^3 \right)
\nonumber \\
& & \quad \quad + \, 12 \left(1 + 4 \rho^2 - 16 \rho^3 - 10 \rho^4 \right) 
\log \left(\frac{1 + \sqrt{1 - 4 \rho^{\phantom{\! 1}}}}
{1 - \sqrt{1 - 4 \rho^{\phantom{\! 1}}}} \right) \Biggr] 
\nonumber \\
& & + \, 8 \, \left[ {\cal M}_{112} (\rho, \eta) - \sqrt{1 - 4 \rho} \, \log( \eta) \right]\Bigr|_{\eta \to 0}
\nonumber \\
& & + \, 8 \, \sqrt{1 - 4 \rho} \, \log \left(\frac{\mu_0^2}{m_b^2}\right),  
\label{eq:R12ccs} \\ 
{\cal C}_{\rho_D, 22}^{(c \bar c s)} & = &
\frac{2}{3} \Biggl[ \sqrt{1 - 4 \rho} \left(-3 + 22 \rho - 34 \rho^2 - 60 \rho^3 \right)
\nonumber \\
& & \quad \quad  - \, 24 \rho \left(1 + \rho + 2 \rho^2 + 5 \rho^3 \right) 
\log \left(\frac{1 + \sqrt{1 - 4 \rho^{\phantom{\! 1}}}}
{1 - \sqrt{1 - 4 \rho^{\phantom{\! 1}}}} \right) \Biggr] 
\nonumber \\
& & + \, 8 \, \left[ {\cal M}_{112} (\rho, \eta) - \sqrt{1 - 4 \rho} \, \log (\eta) \right]\Bigr|_{\eta \to 0}
\nonumber \\
& & + \, 8 \, \sqrt{1 - 4 \rho} \, \log \left(\frac{\mu_0^2}{m_b^2}\right).
\label{eq:R22ccs}
\end{eqnarray}
The dimensionless parameter $\eta = m_q^2/m_b^2$ and the master integral ${\cal M}_{112}$ is defined as
\footnote{ An explicit analytic expression for $\left[ {\cal M}_{112} (\rho, \eta) - \sqrt{1^{\! \! \! \phantom 1} - 4 \rho} \, \log (\eta) \right]\Bigr|_{\eta \to 0}$ 
has been found in Ref.~\cite{Siegen}. }

\begin{equation}
{\cal M}_{112} (\rho, \eta) = 
- \! \! \! \int\limits_{(\sqrt \rho + \sqrt \eta)^2}^{(1 - \sqrt{\rho})^2} \! \! \! d t \,
\frac{\left(t^2 - 2 (1 + \rho) t + (1 - \rho)^2 \right) (t - \eta + \rho)}{t 
\sqrt{\left(t^2 - 2 (1 + \rho) t + (1 - \rho)^2 \right) 
\left(t^2 - 2 t (\eta +\rho) + (\eta -\rho )^2\right)}}.
\end{equation}
Note, that because of $m_d = m_s = 0$ the following relations hold: 
$$
{\cal C}_{\rho_D, nm}^{(c \bar u d)} = {\cal C}_{\rho_D, nm}^{(c \bar u s)},
 \quad
{\cal C}_{\rho_D, nm}^{(c \bar c s)} = {\cal C}_{\rho_D, nm}^{(c \bar c d)},
\quad 
{\cal C}_{\rho_D, nm}^{(u \bar u d)} = {\cal C}_{\rho_D, nm}^{(u \bar u s)},
 \quad
{\cal C}_{\rho_D, nm}^{(u \bar c s)} = {\cal C}_{\rho_D, nm}^{(u \bar c d)}.
$$
The relative effect of the Darwin term with respect to the corresponding partonic-level contribution is given by  
\begin{equation}
\Delta_{\rho_D,nm}^{(q_1 \bar q_2 q_3)} = 
\frac{{\cal C}_{\rho_D,nm}^{(q_1 \bar q_2 q_3)}}{{\cal C}_0^{(q_1 \bar q_2 q_3)}} \, \frac{\rho_D^3 }{m_b^3} \, .
\end{equation}
In Fig.~\ref{fig:Deltas}, these ratios are plotted as functions of $\rho$ for all the colour structures and the four modes,
using for reference the values $\mu_0 = m_b$, $m_b=4.5$ GeV and $\rho_D^3 = 0.2$ GeV$^3$. Fig. \ref{fig:Deltas-sum}
shows the total relative contribution for each mode, namely
\begin{equation}
\Delta_{\rho_D}^{(q_1 \bar q_2 q_3)} =  \frac{ 3 \, C_1^2 \, {\cal C}_{\rho_D, \, 11}^{(q_1 \bar q_2 q_3)} 
+ 2 \, C_1  C_2 \, {\cal C}_{\rho_D,\, 12}^{(q_1 \bar q_2 q_3)} 
+ 3 \, C_2^2 \, {\cal C}_{\rho_D,\, 22}^{(q_1 \bar q_2 q_3) } }
      {\left( 3 \, C_1^2  + 2 \,  C_1 C_2 +3 \, C_2^2 \right) \,  {\cal C}_0^{(q_1 \bar q_2 q_3)}  }\,  \frac{\rho_D^3}{m_b^3} \, .
\end{equation}
As one can see, the Darwin operator leads to sizeable corrections of the order $1-7\, \% $ (for $\rho = 0.05$)
to the $b \to q_1 \bar q_2 q_3$ decay width.

\noindent      
A final comment about the effect of the constant terms $b_i$ in Eq.~(\ref{eq:ME-OT1})
for the determination of the coefficients ${\cal C}_{\rho_D, nm}^{(q_1 \bar q_2 q_3)}$. 
We have already emphasised that their values depend on the choice of the four-quark operators basis. 
Consider as an example the coefficient ${\cal C}_{\rho_D ,12}^{(c \bar u d)}$. In our basis given by Eqs.~(\ref{eq:4q-operators-O12}) and (\ref{eq:4q-operators-O34}), 
${\cal C}_{\rho_D ,12}^{(c \bar u d)} = - 29.0$ for the reference value $\rho = 0.05$. 
If~instead of Eq.~(\ref{eq:4q-operators-O12}) we had considered
the following operators (as it was chosen in Ref.~\cite{Siegen}):
\begin{equation}
\tilde {\cal O}_{6,1}^{(u)} = 
(\bar b_v^{\, i} \Gamma^\sigma \gamma^\mu \Gamma^\rho u^i)
(\bar u^j \Gamma_\sigma \gamma_\mu \Gamma_\rho b_v^{\, j}),
\qquad
\tilde {\cal O}_{6,2}^{(u)} = 
(\bar b_v^{\, i} \Gamma^\sigma \slashed v \Gamma^\rho u^i)
(\bar u^j \Gamma_\sigma \slashed v \Gamma_\rho b_v^{\,j}),
\end{equation} 
then in Eq.~(\ref{eq:ME-OT1}) we would have obtained
$a_1 = 8, \, $
$b_1 = -5/4$ and 
$a_2 = 2, \,$ 
$ b_2 = -3/2$, 
leading to 
${\cal C}_{\rho_D ,12}^{(c \bar u d)} = - 24.0$ (for the same value of $\rho $). This shift of $\sim 17 \%$ should be compensated (up to higher orders) by the different value of the matrix element of the operators defined in these two bases.

\begin{table}[t]\centering
\begin{tabular}{|c|c|c|c||c|c|c|}
\hline
  & ${\cal C}_{G, 11}^{(q_1 \bar q_2 q_3)}$ 
  & ${\cal C}_{G, 12}^{(q_1 \bar q_2 q_3)}$ 
  & ${\cal C}_{G, 22}^{(q_1 \bar q_2 q_3)}$ 
  & ${\cal C}_{\rho_D, 11}^{(q_1 \bar q_2 q_3)}$ 
  & ${\cal C}_{\rho_D, 12}^{(q_1 \bar q_2 q_3)}$ 
  & ${\cal C}_{\rho_D, 22}^{(q_1 \bar q_2 q_3)}$ \\
\hline 
$ b \to u \bar u d $ 
& $-1.5$ & $-9.5$ & $-1.5$
& $6$ & $-11.3$ & $6$  \\
$ b \to u \bar c s $ 
& $-1.28$ & $-6.23$ & $-1.28$
& $6.84$ & $29.1$ & $6.84$  \\
$ b \to c \bar u d $ 
& $-1.28$ & $-8.14$ & $-1.28$
& $-13.2$ & $-29.0$ & $6.84$  \\
$ b \to c \bar c s $ 
& $-1.05$ & $-4.99$ & $-1.05$
& $-11.5$ & $-9.02$ & $7.74$  \\
\hline
\end{tabular}
\caption{Numerical values of the coefficients 
${\cal C}_{G, nm}^{(q_1 \bar q_2 q_3)}$ and ${\cal C}_{\rho_D, nm}^{(q_1 \bar q_2 q_3)}$
for $\rho = 0.05$ and $\mu_0 = m_b$.}
\label{tab:CrhoD-values}
\end{table}

\section{Discussion and conclusion}
\label{sec:discussion}
This work presents the first determination of the Darwin term contribution to the non-leptonic decay $b \to q_1 \bar q_2 q_3$.
Using the expansion of the quark propagator in the external gluon field together with the standard technique of the HQE allows
to obtain a systematic expansion in inverse powers of the heavy quark mass $m_b$. At order $1/m_b^3$ operator mixing ensures
that the IR divergences arising from the expansion of the light quark propagators cancel, introducing though scale-dependence
in the coefficient of $\rho_D^3$, with the divergent $\log(m_q^2/m_b^2)$ being in fact replaced by
$\log(\mu_0^2/m_b^2)$~\footnote{Up to a finite polynomial in $\rho$, depending on the specific four-quark operator basis adopted.}.
Preliminary numerical analysis reveals that this contribution is sizeable.
For illustration purposes, we show the total non-leptonic decay width of the $B^+\! $ and $ B_d$ mesons
up to dimension-six and just at LO-QCD. 
Inputs for the quark masses and the HQE parameters are taken from Refs.~\cite{Alberti:2014yda, Kirk:2017juj, King:2020} and we set $\mu_1 = \mu_0 = m_b$:
\begin{eqnarray}
\Gamma_{\rm NL} (B) & = & 
\Gamma_3 
\Biggl[
    1 -  0.50  \frac{\mu_\pi^2}{m_b^2}
      -  0.44 \frac{\mu_G^2}{m_b^2}
      - 20.25 \frac{\rho_D^3}{m_b^3}
      + \underbrace{. \, . \, .}_{\rm 4q-contr.}
\Biggr]
\nonumber \\
& =& 
\Gamma_3 \biggl[1 - \underbrace{0.0112}_{\mu_\pi^2}  - \underbrace{0.0071}_{\mu_G^2} 
- \underbrace{0.0373}_{\rho_D^3}  - \underbrace{0.0029}_{\tau_i^{(q)}} \, 
\underbrace{\phantom{A \! \! \!}^{\displaystyle { - \, 0.1023} 
\,  (B^+)}_{\displaystyle { + \, 0.0148} \,  (B_d)}}_{{\cal B}_i^{(q)}} \, \biggr],
\label{eq:Gamma-NL-num}
\end{eqnarray} 
where
\begin{equation}
\Gamma_3 = \left( 3 \, C_1^2  + 2 \,  C_1 C_2 +3 \, C_2^2 \right) \! \! \! 
\sum_{\{q_1, q_2, q_3\}} \! \! \! \Gamma_0^{(q_1 \bar q_2 q_3)} \, {\cal C}_0^{(q_1 \bar q_2 q_3)},
\end{equation}
and $\Gamma_0^{(q_1 \bar q_2 q_3)}$ is given by Eq.~(\ref{eq:Gamma0}). 
In Eq.~(\ref{eq:Gamma-NL-num}) the effect of the different non-perturbative parameters is shown separately.
We find that the new contribution due to the Darwin operator 
is significant, for the $B_d$ meson it is even larger than the dimension-five and the weak-exchange contributions while for the $B^+$ meson
the Pauli interference term still gives the dominant correction.
To understand this better,
in Table~\ref{tab:CrhoD-values}
 we compare 
 the coefficients of the Darwin and of the chromo-magnetic operators\,\footnote{For simplicity, we do not show the coefficient of $\mu_\pi^2$ being just $ (- 1/2) \, {\cal C}_0^{(q_1 \bar q_2 q_3)} $ for all three $\Delta B = 1 $ operator combinations.} for different modes.
 As already stated above, at order~$1/m_b^2$, both diagrams in Fig.~\ref{fig:O1-O1}
  vanish and the coefficient of the kinetic operator as well as ${\cal C}_{G, 11}^{(q_1 \bar q_2 q_3)}$ and ${\cal C}_{G, 22}^{(q_1 \bar q_2 q_3)}$
  receive only the small contribution coming from the expansion of the matrix element of dimension-three, c.f.   Eq.~(\ref{ME1}). 
Hence the 
dimension-five contribution stemming from the  $Q_1 \otimes Q_1$ combination 
with large $3 \, C_1^2$ is actually suppressed.  
A further pronounced source of cancellation lies in the combination $3 \,  C_1^2 \, {\cal C}_{G, 11}^{(q_1 \bar q_2 q_3)} + 2 \, C_1 C_2 \, {\cal C}_{G, 12}^{(q_1 \bar q_2 q_3)} $ because of the negative value of $C_2$
and of the hierarchy
$|C_1/C_2| \approx {\cal C}_{G, 12}^{(q_1 \bar q_2 q_3)}/{\cal C}_{G, 11}^{(q_1 \bar q_2 q_3)}$. No cancellations are present at order~$1/m_b^3$ and this explains why, despite the power suppression, the Darwin contribution is larger than the one of dimension-five. 

This result does not come as completely new, in fact it was already known in the well studied inclusive semi-leptonic decays, that the Darwin term constitutes a sizeable correction to the decay rate.
For example, in the inclusive semil-leptonic 
$B \to X_c \ell \bar \nu_\ell$ transition (with $\ell = e, \mu$) the following hierarchy (up to $1/m_b^4$ and at LO-QCD) is found (see e.g. Ref.~\cite{Dassinger:2006md})
\begin{eqnarray}
\Gamma (B \to X_c \ell \bar \nu_\ell) 
& \!\! = \!\! & 
\Gamma_3^{(c \ell \bar \nu)}
\Biggl[
1 - 0.5\frac{\mu^2_\pi}{m_b^2} 
- 1.9 \frac{\mu_G^2}{m_b^2} 
- 19.1 \frac{\rho_D^3}{m_b^3}
- 20.5 \frac{s_1}{m_b^4}
+ 4.0 \frac{s_2}{m_b^4}
- 13.5 \frac{s_3}{m_b^4}
- 9.5 \frac{s_5}{m_b^4}
\Biggr]
\nonumber \\
& \!\! = \!\! & 
\Gamma_3^{(c \ell \bar \nu)} 
\Biggl[
1 - \underbrace{0.043}_{\rm dim-5} - 
\underbrace{0.029}_{\rm dim-6} + 
\underbrace{0.003}_{\rm dim-7}
\Biggr],
\label{eq:SL-width-num-res}
\end{eqnarray}
where $\Gamma_3^{(c \ell \bar \nu)} = \Gamma_0^{(c \ell \bar \nu)} {\cal C}_0^{(c \ell \bar \nu)} $ 
and the numerical values of parameters are taken from Ref.~\cite{Gambino:2016jkc}. 

The size of the coefficients in Eq.~(\ref{eq:SL-width-num-res}) confirms the suppression 
of the dimension-five contribution but does not seem to point towards an abnormal enhancement of the Darwin operator,
whose coefficient is of similar size as the ones of dimension-seven. Note also that the second source of cancellation discussed above is now absent and the dimension-five contribution is larger than in the non-leptonic case. Finally, from Eq.~(\ref{eq:SL-width-num-res})
it does not appear to be any problem with the convergence of the~HQE.

To conclude,
the inclusion of the Darwin operator contribution could lead to a sizeable modification of the theory prediction for the lifetime ratio 
$\tau (B_s) / \tau (B_d)$ compared to the high experimental precision, 
which reaches the accuracy of the four per mille level. An updated theoretical analysis of  $\tau (B_s) / \tau (B_d)$ will be performed once the $SU(3)_F$ violation ratio of the Bag parameters and of the matrix elements of the Darwin operator will become available.

\section*{Acknowledgements}
We would like to thank Thomas Mannel, Daniel Moreno and Alexei Pivovarov for the very valuable discussions and for sharing their results \cite{Siegen}
before publication. These are in perfect agreement with ours and details of the difference in the theoretical methods used can be found in Ref.~\cite{Siegen}.
Moreover we are also grateful to
Simon Badger, Ekta Chaubey, Danny King, Robin Marzucca, Thomas Rauh, Johannes Schlenk and  Keri Vos
for the helpful discussions.
This work was supported by the STFC grant of the IPPP.

\begin{appendix}
\section{Expansion of the quark propagator in the external gluon field}
\label{sec:propagator}

Following Refs.~\cite{Blok:1992hw, Novikov:1983gd},
soft gluon interactions of the quark field can be accounted using the background field method. 
Assuming the $b$ quark embedded in a weakly changing gluon field
allows to systematically expand the quark propagator as a series in the gluon field strength tensor
$G_{\mu \nu} = -i \,[i D_\mu, i D_\nu]$. 
In the case of a massive quark the expansion of the propagator up to terms linear in $G_{\mu\nu}$ is given in Ref.~\cite{Blok:1992hw}, while an expression including 
terms proportional to $D_\rho G_{\mu\nu} = - [i D_\rho, [i D_\mu, i D_\nu]]$ 
can be found in Ref.~\cite{Blok:1994cd} \footnote{
Note that their expression is not complete as it only contains terms relevant for the computation, namely with odd number 
of gamma-matrices.}.
To compute the propagator, one starts from the Green-function equation,
\begin{equation}
(i \slashed \partial + \slashed A (x) - m) S (x, y) = \delta^{(4)} (x-y),
\end{equation}
which admits a solution in form of the perturbative series
\begin{equation}
i S (x, y) = i S^{(0)}(x-y) + i S^{(1)}(x,y) +  \ldots 
\end{equation}
where $S^{(0)}(x-y)$ is the free-quark propagator and $S^{(1)}(x,y)$ the first order correction:
\begin{equation}
i S^{(1)}(x,y) =\int d^4 z \,\,  i S^{(0)} (x - z)\,\, i \slashed A (z)\,\, i S^{(0)} (z - y).
\label{eq.quark-propagator-definition}
\end{equation}
Using the the Fock-Schwinger gauge i.e. $x^\mu  A_\mu(x) = 0$, the gluon field is expressible directly
in terms of the gluon field strength tensor, see Ref.~\cite{Novikov:1983gd} for a detailed derivation: 
\begin{equation}
A_\mu^a (z) = \int_0^1 \! \! d \alpha \, \alpha z^\rho \, G^a_{\rho\mu}(\alpha z).
\end{equation}
Expanding $G_{\rho \mu}(\alpha z)$ around $z = 0$ 
 and taking into account that 
in the Fock-Schwinger gauge $z^\mu \partial_\mu =~z^\mu D_\mu$, yields:
\begin{equation}
A_\mu^a (z) = 
\frac{1}{2} z^\rho G_{\rho\mu}^a(0) + 
\frac{1}{3} z^\nu z^\rho D_\nu G_{\rho\mu}^a(0) + 
{\cal O} (D D G).
\end{equation}
Substituting the previous expression in Eq.~(\ref{eq.quark-propagator-definition}) and setting $y = 0$ 
one obtains
\begin{equation}
S(x,0) = \int \frac{d^4 k}{(2 \pi)^4} \, e^{- i k x} \,  S(k),
\label{eq:quark-propagator}
\end{equation}
where the quark propagator in momentum space reads
\begin{eqnarray}
S(k) & = & 
\frac{\slashed{k} + m}{k^2-m^2} + 
\frac{1}{2 (k^2-m^2)^2} \left( - m  G_{\rho\mu} \sigma^{\rho\mu} +
2 \, \tilde G_{\rho\mu} \, k^\rho \gamma^\mu \gamma^5 \right) 
\nonumber \\
& & 
+ \, \frac{2}{3} \frac{D_\rho G^{\rho\mu}}{(k^2-m^2)^2} 
  \left( \gamma_\mu - \frac{(\slashed{k} + 2 m) \, k_\mu}{k^2-m^2}\right) 
- \frac{2}{3} \frac{D_\nu G_{\rho\mu} \, k^\nu k^\rho \gamma^\mu}{(k^2-m^2)^3}  
\nonumber\\
& & 
+ \, \frac{2}{3} \frac{m \, D_\nu G_{\rho\mu}}{(k^2-m^2)^3} 
\left( \gamma^\rho \gamma^\mu k^\nu - \gamma^\mu \gamma^\nu k^\rho \right)  
+ 2 i \, \frac{D_\nu \tilde G_{\rho \mu} \, k^\nu k^\rho \gamma^\mu \gamma^5}{(k^2-m^2)^3} 
+ \ldots.
\label{eq:quark-propagator-momentum}
\end{eqnarray}
In the above, $\tilde G_{\rho\mu}= (1/2) \epsilon_{\rho \mu \sigma \eta}G^{\sigma \eta}$ and $\epsilon_{\rho\mu\sigma\eta}$ is the Levi-Civita tensor, while the ellipsis stands for terms with higher derivatives as well as higher powers of $G_{\mu \nu}$. 
In the limit $m \rightarrow 0$, Eq.~(\ref{eq:quark-propagator-momentum}) correctly reproduces the
massless expression given in Ref.~\cite{Novikov:1983gd}. 
Finally, we emphasise that the Fock-Schwinger gauge breaks explicitly the translation invariance 
of $S(x,y)$, namely:
$$
S(0,x) = \int \! \! \frac{d^4 k}{(2 \pi)^4} \, e^{ i k x} \, \tilde S(k), \quad \quad \tilde S(k) \neq S(k),
$$
with
\begin{eqnarray}
\tilde S(k) & = & 
\frac{\slashed{k} + m}{k^2-m^2} + 
\frac{1}{2 (k^2-m^2)^2} \left( - m  G_{\rho\mu} \sigma^{\rho\mu} +
2 \, \tilde G_{\rho\mu} \, k^\rho \gamma^\mu \gamma^5 \right) 
\nonumber \\
& & 
+ \, \frac{2}{3} \frac{D_\rho G^{\rho\mu}}{(k^2-m^2)^2} 
  \left( \gamma_\mu - \frac{\slashed{k}  \, k_\mu}{k^2-m^2}\right) 
- \frac{2}{3} \frac{D_\nu G_{\rho\mu} \, k^\nu k^\rho \gamma^\mu}{(k^2-m^2)^3}  
\nonumber\\
& & 
- \, \frac{2}{3} \frac{m \, D_\nu G_{\rho\mu}}{(k^2-m^2)^3} 
\left( \gamma^\rho \gamma^\mu k^\nu - \gamma^\mu \gamma^\nu k^\rho \right)  
- 2 i \, \frac{D_\nu \tilde G_{\rho \mu} \, k^\nu k^\rho \gamma^\mu \gamma^5}{(k^2-m^2)^3} 
+ \ldots.
\label{eq:quark-propagator-momentum-tilde}
\end{eqnarray}
For completeness, we also present the equivalent representation of Eq.~(\ref{eq:quark-propagator}) in coordinate space
\begin{eqnarray}
S(x,0) & = & 
- \frac{i}{4\pi^2} \frac{m^2 K_1 (m \sqrt{-x^2})}{\sqrt{-x^2}} 
- \frac{1}{4 \pi^2} \frac{m^2 \slashed x \, K_2 (m \sqrt{-x^2})}{x^2} 
\nonumber \\
&-& \frac{\tilde G_{\alpha \beta}}{8 \pi^2} x^\alpha \gamma^\beta \gamma_5
    \frac{m K_1 (m \sqrt{-x^2})}{\sqrt{-x^2}} 
 - i  \frac{G_{\alpha \beta}}{16 \pi^2} \sigma^{\alpha \beta} m K_0 (m \sqrt{-x^2}) 
\nonumber \\
&+& \frac{i}{24 \pi^2} D_\alpha G^{\alpha \beta} \gamma_\beta K_0 (m \sqrt{-x^2})
  - \frac{i}{48 \pi^2} D_\alpha G^{\alpha \beta} x_\beta \, \slashed x \,
    \frac{m K_1 (m \sqrt{-x^2})}{\sqrt{-x^2}}  
\nonumber \\
&-& \frac{1}{24 \pi^2} D_{\alpha} G^{\alpha \beta} x_\beta \, m K_0 (m \sqrt{-x^2})
  - \frac{i}{48 \pi^2} D^{\alpha} G^{\beta \rho} \gamma_\rho x_\alpha x_\beta \, 
    \frac{m K_1 (m \sqrt{-x^2})}{\sqrt{-x^2}}
\nonumber \\
&-& \frac{1}{16 \pi^2} D^{\alpha} \tilde G^{\beta \rho} \gamma_\rho \gamma_5 x_\alpha x_\beta \, 
    \frac{m K_1 (m \sqrt{-x^2})}{\sqrt{-x^2}}
 -  \frac{1}{48 \pi^2} D^{\alpha} G^{\beta \rho} \gamma_\rho \gamma_\beta 
    x_\alpha \, m K_0 (m \sqrt{-x^2})
\nonumber \\
&-& \frac{1}{48 \pi^2} D^{\alpha} G^{\beta \rho} \gamma_\rho \gamma_\alpha x_\beta \, 
    m K_0 (m \sqrt{-x^2}) + {\cal O} (D D G),
\label{eq:propagator}
\end{eqnarray}
where $K_{0,1,2} (z)$ are the modified Bessel function of the second kind.
\section{Complementary material to Section \ref{sec:logs}}
\label{App:Mixing}
The divergent coefficients ${\cal D}_{nm}^{( q_1 \bar q_2 q_3)}$ in Eq.~(\ref{eq:C-rho-D-lim}) read:
\begin{eqnarray}
{\cal D}_{11}^{(u \bar u d)}  & = & 8\,  \log \left( \frac{m_u^2}{m_b^2} \right),
\nonumber \\
{\cal D}_{12}^{( u \bar u d)} & = &8 \,\left[ \log \left( \frac{m_d^2}{m_b^2} \right) - \log \left( \frac{m_u^2}{m_b^2} \right) \right], 
\nonumber \\
 {\cal D}_{22}^{(u \bar u d)} & = & 8 \,  \log \left( \frac{m_d^2}{m_b^2} \right), 
\\[5mm]
{\cal D}_{11}^{(u  \bar c s)} & = & 8 \, (1- \rho)^2 \, (1 + \rho ) \log \left( \frac{m_u^2}{m_b^2} \right),
\label{eq:Du} 
\nonumber \\ 
{\cal D}_{12}^{(u  \bar c s)} & = & 8 \, (1- \rho)^2 \, (1 + \rho )  \left[ \log \left( \frac{m_u^2}{m_b^2} \right) + \log \left( \frac{m_s^2}{m_b^2} \right)\right],
\nonumber \\
{\cal D}_{22}^{(u \bar c s)} & = & 
8 \, (1- \rho)^2 \, (1 + \rho ) \log \left( \frac{m_s^2}{m_b^2} \right),
\\[5mm]
{\cal D}_{12}^{(c \bar u d)} & = & - 16 \, (1 - \rho)^2             \log \left( \frac{m_u^2}{m_b^2} \right) + 
		        	  8 \, (1 - \rho)^2 (1 + \rho)  \log \left( \frac{m_d^2}{m_b^2} \right), 
\nonumber \\
{\cal D}_{22}^{(c \bar u d)} & = &  8 \, (1 - \rho)^2 (1 + \rho) \log \left( \frac{m_d^2}{m_b^2} \right), 
\label{eq:Dcud}
\\[5mm]
{\cal D}_{12}^{(c \bar c s)} & = &  
8 \, \sqrt{1 - 4 \rho} \, \log \left( \frac{m_s^2}{m_b^2} \right),
\nonumber \\
{\cal D}_{22}^{(c \bar c s)} & = & 
8 \, \sqrt{1 - 4 \rho} \, \log \left( \frac{m_s^2}{m_b^2} \right).
\end{eqnarray}

The discontinuity of the WA, PI and WE diagrams in Fig.~\ref{fig:PI-and-WE} at LO-QCD, respectively is
\begin{eqnarray}
\Gamma^{\rm WA} 
& = & 
\frac{\Gamma_0}{2 m_B} \frac{16 \pi^2}{m_b^3}  \sqrt{\lambda(1, z_2, z_3)} \, \Biggl\{ \left(N_c C_1^2 + 2 \, C_1 C_2 \right)
\nonumber \\
& & 
\times \,  \biggl[  \left( (z_2 - z_3)^2 + z_2 + z_3 - 2 \right) \tilde {\cal O}_{6,1}^{(q_1)} 
\nonumber \\
& & 
\quad - \, 2 \left( 2 (z_2 - z_3)^2 - 1 - z_2 -  z_3 \right) \tilde {\cal O}_{6,2}^{(q_1)} \biggl] 
\nonumber \\
 & & 
 + \, C_2^2 \, \biggl[  \left( (z_2 - z_3)^2 + z_2 + z_3 - 2 \right) \tilde {\cal O}_{6,3}^{(q_1)} 
\nonumber \\
& & 
\qquad - \, 2  \left( 2 (z_2 - z_3)^2 - 1 - z_2 -  z_3 \right) \tilde {\cal O}_{6,4}^{(q_1)} \biggl]
\Biggr\},
\label{eq:WA} 
\end{eqnarray}
\begin{eqnarray}
\Gamma^{\rm PI}  
& = &
\frac{\Gamma_0}{2 m_B} \frac{96 \pi^2}{m_b^3}  \sqrt{\lambda(1,z_1,z_3)} \,  \left(1-z_1-z_3 \right)\
\nonumber \\
& &
\biggl[2 \, C_1 C_2\, \tilde {\cal O}_{6,1}^{(q_2)} + \left( C_1^2 +  C_2^2 \right) \, \tilde {\cal O}_{6,3}^{(q_2)} \biggr],
\label{eq:PI} 
\\[5mm]
\Gamma^{\rm WE} 
& = & 
\frac{\Gamma_0}{2 m_B} \frac{16 \pi^2}{m_b^3}  \sqrt{\lambda(1,z_1,z_2)} \, \Biggl\{ \left(2 \, C_1 C_2 + N_c C_2^2 \right)
\nonumber \\
& & 
\times \,  \biggl[  \left( (z_1 - z_2)^2 + z_1 + z_2 - 2 \right) \tilde {\cal O}_{6,1}^{(q_3)} 
\nonumber \\
& & 
\quad - \, 2 \left( 2 (z_1 - z_2)^2 - 1 - z_1 -  z_2 \right) \tilde {\cal O}_{6,2}^{(q_3)} \biggl] 
\nonumber \\
 & & 
 + \, C_1^2 \, \biggl[  \left( (z_1 - z_2)^2 + z_1 + z_2 - 2 \right) \tilde  {\cal O}_{6,3}^{(q_3)} 
\nonumber \\
& & 
\qquad - \, 2  \left( 2 (z_1 - z_2)^2 - 1 - z_1 -  z_2 \right)  \tilde {\cal O}_{6,4}^{(q_3)} \biggl]
\Biggr\},
\label{eq:WE} 
\end{eqnarray}
where $z_i = m_{q_i}^2/m_b^2$ and the four-quark operators $\tilde {\cal O}_{6,i}^{(q)}$ are
defined in Eqs.~(\ref{eq:4q-operators-O12}), (\ref{eq:4q-operators-O34}).
Note that, since we set $m_u = m_d = m_s = 0$, in our case $z_i$ can be either equal to $\rho $ or 0.
\section{Contribution of four-quark operators at order~\boldmath $1/m_b^3$ }
\label{App:Xs}
The four-quark operators contribution to the non-leptonic decay $b \to q_1 \bar q_2 q_3 $ at order $1/m_b^3$ and at LO-QCD is written in the following form
\begin{eqnarray}
\Gamma^{(\rm 4 q)}_{\rm NL} (B) & = &
\frac{\Gamma_0}{2 m_B} \,  \Biggl( 3 \, C_1^2 \, {\cal P}_{11}^{(q_1 \bar q_2 q_3)}
+ 2 \, C_1  C_2 \, {\cal P}_{12}^{(q_1 \bar q_2 q_3)} 
+ 3 \, C_2^2 \, {\cal P}_{ 22}^{(q_1 \bar q_2 q_3) } \Biggr),
\label{eq:Gamma-NL-final-res}
\end{eqnarray}
where
\begin{eqnarray} 
P_{11}^{(u \bar u d)} & = & 
\frac{32 \pi^2}{m_b^3} 
\left[
- \langle \tilde{\cal O}_{6,1}^{(u)}  \rangle + \langle \tilde{\cal O}_{6, 2}^{(u)} \rangle + \langle \tilde{\cal O}_{6,3}^{(u)} \rangle - \frac{ \langle \tilde{\cal O}_{6,3}^{(d)} \rangle - \langle \tilde{\cal O}_{6,4}^{(d)} \rangle }{3}
\right],
\nonumber \\
P_{12}^{(u \bar u d)} & = & 
\frac{ 32 \pi^2 }{m_b^3} 
\left[
2 \, \langle \tilde{\cal O}_{6,1}^{(u)} \rangle + \langle \tilde{\cal O}_{6,2}^{(u)} \rangle - \langle \tilde{\cal O}_{6,1}^{(d)} \rangle + \langle \tilde{\cal O}_{6,2}^{(d)} \rangle  
\right], \\
P_{22}^{(u \bar u d)} & = & 
\frac{32 \pi^2}{m_b^3} 
\left[
- \langle \tilde{\cal O}_{6,1}^{(d)} \rangle + \langle \tilde{\cal O}_{6,2}^{(d)} \rangle + \frac{2 \langle \tilde{\cal O}_{6,3}^{(u)} \rangle + \langle \tilde{\cal O}_{6,4}^{(u)} \rangle}{3}
\right], 
\nonumber
\end{eqnarray}
\begin{eqnarray}
P_{11}^{(u \bar c s)} & = &  
\frac{32 \pi^2}{m_b^3} (1 - \rho)^2
\biggl[
(1 + 2 \rho) \, 
\left(
\langle \tilde{\cal O}_{6,2}^{(u)} \rangle + 
\frac{1}{3} \langle \tilde{\cal O}_{6,4}^{(s)} \rangle 
\right) 
- 
\left(1 + \frac{\rho}{2} \right) 
\left(
\langle \tilde{\cal O}_{6,1}^{(u)} \rangle 
+ \frac{1}{3} \langle \tilde{\cal O}_{6,3}^{(s)} \rangle
\right)
\biggr],
\nonumber \\
P_{12}^{(u \bar c s)} & = & 
\frac{32 \pi^2}{m_b^3} (1 - \rho)^2
\biggl[
(1 + 2 \rho) \left( \langle \tilde{\cal O}_{6,2}^{(u)} \rangle + \langle \tilde{\cal O}_{6.2}^{(s)} \rangle \right) 
- \left(1 + \frac{\rho}{2} \right) \left(\langle \tilde{\cal O}_{6,1}^{(u)} \rangle + \langle \tilde{\cal O}_{6,1}^{(s)} \rangle \right)
\biggr], 
\\
P_{22}^{(u \bar c s)} & = & 
\frac{32 \pi^2}{m_b^3} (1 - \rho)^2
\left[
(1 + 2 \rho) \left(\langle \tilde{\cal O}_{6,2}^{(s)}  \rangle + 
\frac{1}{3} \langle \tilde{\cal O}_{6,4}^{(u)} \rangle \right) 
-\left(1 + \frac{\rho}{2} \right) \left(\langle \tilde{\cal O}_{6,1}^{(s)} \rangle  + \frac{1}{3} \langle \tilde{\cal O}_{6,3}^{(u)} \rangle \right)
\right], 
\nonumber 
\end{eqnarray}
\begin{eqnarray}
P_{11}^{(c \bar u d)} & = & 
\, \frac{32 \pi^2 }{3 m_b^3} (1 - \rho)^2
\left[
(1 + 2 \rho) \, \langle \tilde{\cal O}_{6,4}^{(d)} \rangle - \left(1 + \frac{\rho}{2} \right) \langle \tilde{\cal O}_{6,3}^{(d)} \rangle + 3 \,\langle \tilde{\cal O}_{6,3}^{(u)} \rangle 
\right], 
\nonumber \\
P_{12}^{(c \bar u d)} & = & 
\frac{32 \pi^2 }{m_b^3} (1 - \rho)^2
\left[
(1 + 2 \rho) \, \langle \tilde{\cal O}_{6,2}^{(d)} \rangle - \left(1 + \frac{\rho}{2} \right) \langle \tilde{\cal O}_{6,1}^{(d)} \rangle + 3 \, \langle \tilde{\cal O}_{6,1}^{(u)} \rangle
\right], 
\\
P_{22}^{(c \bar u d)} & = & 
\frac{32 \pi^2 }{m_b^3} (1 - \rho)^2
\left[
(1 + 2 \rho) \langle \tilde{\cal O}_{6,2}^{(d)} \rangle - \left(1 + \frac{\rho}{2} \right) \langle \tilde{\cal O}_{6,1}^{(d)} \rangle + \langle \tilde{\cal O}_{6,3}^{(u)} \rangle
\right], 
\nonumber 
\end{eqnarray}
\begin{eqnarray}
P_{11}^{(c \bar c s)} & = & 
\, \frac{32 \pi^2 }{3 m_b^3} \sqrt{1 - 4 \rho}
\left[
(1 + 2 \rho) \, \langle \tilde{\cal O}_{6,4}^{(s)}\rangle  - \left(1 - \rho \right) \langle \tilde{\cal O}_{6,3}^{(s)} \rangle 
\right],
\nonumber \\
P_{12}^{(c \bar c s)} & = & 
\frac{32 \pi^2 }{m_b^3} \sqrt{1 - 4 \rho}
\left[
(1 + 2 \rho) \, \langle \tilde{\cal O}_{6,2}^{(s)} \rangle  - \left(1 - \rho \right) \langle \tilde{\cal O}_{6,1}^{(s)} \rangle 
\right],
\\
P_{22}^{(c \bar c s)} & = & 
\frac{32 \pi^2 }{m_b^3} \sqrt{1 - 4 \rho}
\left[
(1 + 2 \rho) \, \langle \tilde{\cal O}_{6,2}^{(s)} \rangle  - \left(1 - \rho \right) \langle \tilde{\cal O}_{6,1}^{(s)} \rangle
\right].
\nonumber 
\end{eqnarray}
The corresponding expressions for ${\cal P}_{nm}^{(u \bar u s)}$,
${\cal P}_{nm}^{(u \bar c d)}$, ${\cal P}_{nm}^{(c \bar u s)}$, ${\cal P}_{nm}^{(c \bar c d)}$
can be obtained from the above ones by replacing 
$ \tilde{\cal O}_{6,i}^{(d)} \leftrightarrow  \tilde{\cal O}_{6,i}^{(s)}$.
\section{Coefficients of the dimension-three and of the chromo-magnetic operators}
\label{App:C3C5}
Here we present the analytic expressions for the coefficients of the dimension-three and chromo-magnetic operators 
intoduced in Eq.~(\ref{eq:Gamma-NL-res-scheme}). 
They read respectively
\begin{eqnarray}
& & {\cal C}_{0}^{(u \bar u d)} = 1, 
\qquad {\cal C}_{G,11}^{(u \bar u d)} = {\cal C}_{G,22}^{(u \bar u d)}  
= - \frac{3}{2},  \qquad
{\cal C}_{G,12}^{(u \bar u d)}  =  -\frac{19}{2},
\end{eqnarray}
\begin{eqnarray}
{\cal C}_{0}^{(u \bar c s)} 
& = & 
1 - 8 \rho - 12 \rho^2 \log (\rho) + 8 \rho^3 - \rho^4,
\nonumber \\
{\cal C}_{G,11}^{(u \bar c s)} = {\cal C}_{G,22}^{(u \bar c s)}  
& = & 
-\frac{1}{2} \left(3 - 8 \rho + 12 \rho^2 \log( \rho) 
+ 24 \rho ^2 - 24 \rho^3 + 5 \rho^4 \right),
\\
{\cal C}_{G,12}^{(u \bar c s)} & = & 
-\frac{1}{2} 
\left(19 + 16 \rho + 12 \rho (\rho + 4) \, \log(\rho) - 24 \rho ^2 - 16 \rho ^3 
+ 5 \rho^4  \right),
\nonumber
\end{eqnarray}
\begin{eqnarray}
{\cal C}_{0}^{(c \bar u d)} & = & 
1 - 8 \rho - 12 \rho^2 \log (\rho) + 8 \rho^3 - \rho^4,
\nonumber \\
{\cal C}_{G,11}^{(c \bar u d)} = {\cal C}_{G,22}^{(c \bar u d)}  & = & 
-\frac{1}{2} \left(3 - 8 \rho + 12 \rho^2 \log( \rho) + 24 \rho ^2   
- 24 \rho^3 + 5 \rho^4 \right),
\label{eq:C0ud} \\
{\cal C}_{G,12}^{(c \bar u d)} & = & 
-\frac{1}{2} 
\left(19 - 56 \rho + 12 \rho^2 \log(\rho) + 72 \rho ^2 - 40 \rho ^3 +5 \rho^4  \right),
\nonumber
\\[5mm]
{\cal C}_{0}^{(c \bar c s)} & = & 
\sqrt{1 - 4 \rho} \left(1  - 14 \rho - 2 \rho^2 - 12 \rho^3 \right) 
+ 24 \rho^2 (1 - \rho^2)  
\log \left(\frac{1 + \sqrt{1 - 4 \rho^{\phantom{\! 1}}}}
{1 - \sqrt{1 - 4 \rho^{\phantom{\! 1}}}} \right) \! ,
\nonumber \\
\! \! \! \! {\cal C}_{G,11}^{(c \bar c s)} = {\cal C}_{G,22}^{(c \bar c s)} & = & 
- \frac{1}{2} \Biggl[
\sqrt{1 - 4 \rho} \left(3  - 10 \rho + 10 \rho^2 + 60 \rho^3 \right) 
\nonumber \\
& & \qquad - \, 24 \rho^2 (1 - 5 \rho^2)  
\log \left(\frac{1 + \sqrt{1 - 4 \rho^{\phantom{\! 1}}}}
{1 - \sqrt{1 - 4 \rho^{\phantom{\! 1}}}} \right) \Biggr],
\label{eq:CG11-22-cs} \\
{\cal C}_{G, 12}^{(c \bar c s)} & = &
- \frac{1}{2} \Biggl[
\sqrt{1 - 4 \rho} \left(19  - 2 \rho + 58 \rho^2 + 60 \rho^3 \right) 
\nonumber \\
& & \qquad - \, 24 \rho \, (2 + \rho - 4 \rho^2 - 5 \rho^3)  
\log \left(\frac{1 + \sqrt{1 - 4 \rho^{\phantom{\! 1}}}}
{1 - \sqrt{1 - 4 \rho^{\phantom{\! 1}}}} \right) \Biggr].
\nonumber
\end{eqnarray}

\end{appendix}

\bibliographystyle{hieeetr}
\bibliography{References}

\end{document}